\documentclass[a4paper,USenglish,cleveref,autoref,thm-restate]{lipics-v2021}

\hideLIPIcs

\newif\ifEditMode

\EditModetrue

\title{Signals and Spoils: Speculative Oracle Extractable Value in the Era of Cross-Chain Interoperability}

\titlerunning{Speculative Oracle Extractable Value in the Era of Cross-Chain Interoperability}

\author{Hasret Ozan Sevim}{University of Camerino, Italy; Catholic University of Sacred Heart, Italy; INESC-ID, Portugal}{hasretozan.sevim@unicam.it}{}{}
\author{Christof Ferreira Torres}{Instituto Superior Técnico (IST), University of Lisbon, Portugal; INESC-ID, Portugal}{christof.torres@tecnico.ulisboa.pt}{}{}

\authorrunning{H.O. Sevim and C.F. Torres}

\authorrunning{ }


\renewcommand\ccsdesc[2][]{}
\makeatletter
\def\@concepts{}
\makeatother


\usepackage{booktabs}
\usepackage{multirow}
\usepackage{pifont}
\usepackage{makecell}
\usepackage{hyperref}
\usepackage{threeparttable}   
\usepackage{tikz}
\usepackage{float}
\usepackage{adjustbox}
\usepackage{enumitem}
\usetikzlibrary{arrows.meta,shapes.geometric,positioning,fit,backgrounds,calc}
\usepackage[colorinlistoftodos]{todonotes}
\usepackage[table]{xcolor}

\usepackage{caption}
\usepackage{subcaption}
\usepackage{siunitx}
\newcommand{\cmark}{\ding{51}}
\newcommand{\xmark}{\ding{55}}

\nolinenumbers 

\begin{document}
\maketitle

\begin{abstract}

A new form of Maximal Extractable Value (MEV), termed speculative MEV, has emerged across Layer-2 blockchains. Unlike Ethereum mainnet, many Layer-2 systems lack a public mempool, forcing extraction strategies to become probabilistic: searchers emit multiple identical transactions hoping to capture an opportunity first. This generates substantial transaction spam, increasing fees and wasting block space. We investigate speculative Oracle Extractable Value (OEV), a form of MEV associated with liquidating undercollateralized loans via speculative backrunning of oracle price updates. We propose a methodology for detecting speculative liquidations in the wild and apply it across Arbitrum, Base, and Optimism. On October 10, 2025, we identify 64 speculative liquidators on Aave (57\% of all detected liquidators) and 831 successful speculative liquidations (39\% of all successful liquidations across the three chains).
We further examine whether latency differences in oracle price feed updates across blockchains can be exploited for cross-chain OEV. Specifically, we ask whether a searcher can observe oracle updates on one chain and frontrun liquidation opportunities on another. We systematically analyze Chainlink Decentralized Oracle Network (DON) configurations (deviation thresholds, heartbeat intervals, and submitted price observations) across Arbitrum, Base, Ethereum, and Optimism. Our dataset comprises 63 Chainlink feeds, 12,009 price updates, and over 100,000 oracle observations linked to 2,986 Aave liquidations. We show that independent DONs consume largely identical off-chain price data nearly simultaneously yet publish updates at different times, creating statistically predictable cross-chain exploitation windows. We demonstrate that Chainlink updates on Optimism can predict subsequent updates on Arbitrum and Base, enabling speculative cross-chain OEV extraction.
\end{abstract}

\section{Introduction}
\label{sec:intro}

Decentralized Finance (DeFi) lending protocols such as Aave have become a fundamental component of the blockchain ecosystem, enabling users to borrow assets against overcollateralized positions without relying on centralized intermediaries. To preserve protocol solvency, these systems employ automated liquidation mechanisms that allow third parties to repay undercollateralized debt positions in exchange for discounted collateral. As a result, liquidations have evolved into one of the most competitive and profitable domains of Maximal Extractable Value (MEV), generating hundreds of millions of dollars in revenue for specialized searchers and liquidators over recent years \cite{Torres2025}.

Traditionally, liquidation strategies have been viewed as reactive: searchers monitor the blockchain for positions whose health factors (HF) have already fallen below the liquidation threshold and compete to execute the liquidation transaction first. However, the rapid expansion of Layer-2 (L2) ecosystems has fundamentally altered this landscape. Unlike Ethereum mainnet, several prominent L2s, including Arbitrum, Optimism, and Base, do not expose a public mempool. Consequently, conventional priority gas auction mechanisms become substantially less effective, forcing searchers to adopt probabilistic transaction placement strategies \cite{Solmaz2025OptimisticMEV}. In practice, this has led to the emergence of a new form of speculative MEV, in which searchers repeatedly emit transactions in anticipation of future state changes rather than reacting to already materialized opportunities. Such speculative behavior generates significant on-chain transaction spam, as many transactions merely perform state reads and are nevertheless included on-chain, consuming block space and increasing transaction fees.

In this paper, we investigate the speculative Oracle Extractable Value (OEV) arising from DeFi liquidations. More specifically, we study whether liquidators can anticipate oracle price updates before they are published on-chain and strategically position liquidation transactions to capture profitable opportunities immediately after the update occurs. This phenomenon is particularly relevant in oracle-driven lending protocols such as Aave, where liquidation eligibility is directly determined by externally supplied price feeds. Because these price feeds are updated asynchronously across different blockchains, oracle latency itself may become a source of extractable value.
Our work is motivated by an important structural observation regarding cross-chain oracle deployments. Independent Chainlink Decentralized Oracle Networks (DONs) on Ethereum and major L2s consume largely identical off-chain market data from centralized exchanges (CEX) and data aggregators nearly simultaneously, yet publish their price updates on-chain at different times. These timing differences arise from heterogeneous block production intervals, oracle deviation thresholds, heartbeat configurations, and network-specific transaction propagation characteristics. As a consequence, an oracle update observed on one blockchain may reveal statistically predictive information about imminent oracle updates on another blockchain, potentially enabling searchers to speculate on forthcoming liquidation opportunities before they become visible locally.

To systematically study this phenomenon, we collect and analyze a large-scale dataset comprising 2,986 Aave liquidation events, 12,009 Chainlink oracle update events, and more than 100,000 individual DON price observations across Arbitrum, Base, Ethereum, and Optimism on October~10,~2025. We further combine this data with second-level spot prices from Binance to characterize the latency relationships between off-chain market movements, oracle observations, on-chain oracle transmissions, and liquidation execution. Based on this analysis, we formalize the notion of a cross-chain exploitation window and investigate whether these windows can be systematically exploited for speculative liquidation strategies.

Beyond modeling the phenomenon theoretically, we also provide empirical evidence that speculative liquidation behavior is already prevalent in production systems. We introduce a methodology for detecting speculative liquidations in the wild based on transaction placement patterns relative to oracle price updates. Applying our methodology across Arbitrum, Base, and Optimism, we identify 64 speculative liquidators, representing 57\% of all detected liquidators, and 831 successful speculative liquidations, corresponding to 39\% of all detected successful liquidations on Aave during our observation period. Our findings indicate that speculative OEV extraction is not merely theoretical, but already constitutes a substantial component of liquidation activity on modern L2 ecosystems.

Overall, our results reveal a previously underexplored interaction between oracle architectures, cross-chain latency asymmetries, and MEV extraction strategies. The findings demonstrate that independently operated oracle networks may unintentionally leak predictive information across chains, thereby creating systematic opportunities for speculative OEV extraction. More broadly, our study highlights how architectural fragmentation across L2 ecosystems can introduce new classes of MEV and raises important questions regarding oracle design, fairness, transaction spam, and the long-term scalability of DeFi infrastructure.

\noindent
\textbf{Contributions.} Our main contributions are summarized as follows:

\begin{itemize}
\item We provide the first empirical study of speculative OEV in the wild, thereby providing empirical evidence that speculative liquidators already exploit oracle timing asymmetries in production environments.

\item We show that speculative liquidation strategies generate substantial profits but also transaction spam and reverted transaction activity, highlighting previously underexplored scalability and efficiency implications of speculative OEV on Layer-2 systems.

\item We characterize the latency relationships between off-chain market movements, DON observations, oracle transmissions, and liquidation execution across multiple blockchains.

\item We formalize the notion of a cross-chain oracle exploitation window and demonstrate that oracle updates on one blockchain can statistically predict imminent updates on another.
\end{itemize}

\section{Background}
\label{sec:background}

In this section, we provide the necessary background on Chainlink’s price oracle architecture, the liquidation mechanism of Aave, and the distinction between Maximal Extractable Value (MEV) and Oracle Extractable Value (OEV).

\subsection{Chainlink Price Feeds}

Chainlink price feeds operate through the Off-Chain Reporting (OCR) protocol, in which a Decentralized Oracle Network (DON) of independent oracle nodes collectively aggregates off-chain market data and periodically publishes updated prices on-chain. Each oracle node independently retrieves asset prices from external data sources and submits signed price observations to a designated transmitter node. The transmitter aggregates these observations and submits a transaction to the on-chain aggregator contract, which emits a \texttt{NewTransmission} event containing the updated median price and the underlying observations.

The oracle architecture consists primarily of two contracts. The \emph{aggregator contract} receives signed OCR reports and stores the latest oracle price, while the \emph{proxy contract} serves as the consumer-facing interface through which applications retrieve price data via \texttt{latestRoundData()}. Since the aggregator contract emits all oracle update events, it represents the primary source for observing on-chain price update activity.

Oracle updates are triggered under two conditions: (i)~when the asset price deviates from the previously reported on-chain value by more than a predefined \emph{deviation threshold} (DT), or (ii)~when a maximum idle period, referred to as the \emph{heartbeat} (HB), has elapsed without an update. These parameters jointly determine the responsiveness of a price feed and directly influence the timing at which new oracle prices are propagated on-chain. Because DON configurations may differ across blockchains, oracle updates for the same asset can occur at different times on different chains despite relying on largely identical off-chain price information.

\subsection{Aave Liquidation Mechanics}

Aave~V3 determines the solvency of collateralized positions through its \texttt{AaveOracle} contract, which maps each collateral and debt asset to an associated Chainlink price feed. During liquidation execution, the call path proceeds from the liquidator to the Aave pool contract, which queries the market’s address provider and oracle contracts before ultimately retrieving asset prices from the corresponding Chainlink aggregator contracts. Depending on the asset type, Aave supports multiple oracle configurations, including standard Chainlink feeds, derivative-based feeds that compose prices from underlying assets, and stablecoin-specific USD aggregators. 

A position becomes eligible for liquidation once its HF $H$ falls below~1. The health factor is computed as

\begin{equation}
H = \frac{\displaystyle\sum_i \text{collateral}_i \times P_i^{\text{oracle}} \times
\text{LT}_i}{\displaystyle\sum_j \text{debt}_j \times P_j^{\text{oracle}}}
\label{eq:hf}
\end{equation}

where $\text{LT}_i$ denotes the liquidation threshold of collateral asset $i$, and
$P^{\text{oracle}}$ represents the oracle price obtained from Chainlink. Since the health
factor depends directly on on-chain oracle prices, every oracle price update may immediately
change the liquidation status of a position and thereby create a potential liquidation
opportunity.

\subsection{Maximal and Oracle Extractable Value}

Maximal Extractable Value (MEV) refers to the profit obtained by strategically influencing the ordering, insertion, or exclusion of transactions within a block. In decentralized finance, MEV commonly arises through arbitrage, liquidations, and other transaction-ordering opportunities. Traditional MEV strategies are typically reactive, meaning that searchers compete to exploit opportunities that have already appeared on-chain.

Oracle Extractable Value (OEV) is a specialized form of MEV that arises in protocols whose state depends on oracle price updates. In lending protocols such as Aave, liquidation eligibility is directly determined by oracle-reported asset prices. Consequently, oracle updates themselves become economically valuable events that can create profitable liquidation opportunities. Conventional OEV strategies therefore attempt to execute liquidation transactions immediately after an oracle price update is published on-chain.

More recently, Layer-2 blockchains have enabled the emergence of \emph{speculative} MEV and OEV strategies. Rather than reacting to finalized state changes, speculative searchers attempt to predict profitable events before they occur on-chain. In the context of speculative OEV, liquidators anticipate forthcoming oracle price updates and pre-position liquidation transactions to execute within the same block as, or immediately after, the oracle update transaction.

\subsection{Smart Value Recapture}

Chainlink’s Smart Value Recapture (SVR) mechanism introduces a modified oracle update process designed to internalize Oracle Extractable Value (OEV). Instead of publishing oracle updates through the standard \texttt{NewTransmission} process, SVR routes updates through a sealed-bid auction in which searchers compete for the right to execute the associated liquidation opportunity. This allows liquidation-related MEV to be redirected from external searchers toward the protocol ecosystem.

In SVR-enabled Aave~V3 markets, liquidators are effectively selected before the oracle update is published on-chain through a private off-chain relay. Consequently, liquidation transactions are typically executed within the same block as the corresponding oracle update, eliminating most public on-chain competition among liquidators. The resulting auction proceeds are subsequently used for protocol-aligned purposes such as LINK and AAVE token buyback programs. 
SVR was operationally active only on Ethereum during our study period and had not yet been deployed across Layer-2 networks.

\section{Related Work}
\label{sec:related}

Recent literature has extensively examined the accuracy, configuration, and economic implications of Chainlink oracle price feeds. Studies such as \cite{PriceOracle2025}, \cite{Nadler2026}, and \cite{Vakhmyanin2023} analyze how heartbeat intervals, deviation thresholds, and network conditions affect oracle update precision and market efficiency. \cite{PriceOracle2025} compares Chainlink price feeds across multiple blockchains against CEX benchmarks and highlights how oracle configurations create temporary price discrepancies and arbitrage opportunities. Similarly, \cite{Nadler2026} studies the relationship between oracle design parameters and price inaccuracies on Ethereum, while \cite{Vakhmyanin2023} shows that delayed oracle updates during volatile market conditions can enable profitable arbitrage strategies.

Beyond oracle accuracy, several works investigate Oracle Extractable Value (OEV), a specialized form of MEV arising from oracle-driven state changes such as liquidations. Torres et al.~\cite{Torres2025} analyze MEV activity across Ethereum and major rollups, documenting liquidation competition and observing same-block extraction patterns on Arbitrum and Optimism despite the absence of public mempools. \cite{Greene2024} characterizes OEV as an inherent consequence of deviation-threshold oracle mechanisms and propose auction-based mitigations to redistribute extracted value. \cite{Andreoulis2026} empirically studies OEV in Aave V2 and V3 across Ethereum and rollups, showing that liquidation profits are highly concentrated during periods of market stress and dominated by a small number of sophisticated searchers. The study additionally introduces the Oracle Update Rebate Window (OURW) model as a mechanism for redirecting liquidation profits back to protocols, which is currently deployed by Chainlink in the form of SVR. Importantly, \cite{Finkbeiner2025} explicitly identifies cross-chain latency arbitrage in OEV markets as an underexplored research direction.

Related MEV studies further emphasize the role of latency and cross-chain inefficiencies. \cite{Gogol2024} analyzes arbitrage opportunities across Layer-2 networks and demonstrates how gas fees and block production times shape persistent cross-chain price discrepancies. Other works, including \cite{Humphry2024}, \cite{Yang2024}, and \cite{Qin2021}, discuss oracle manipulation risks, liquidation backrunning, and liquidation dynamics in major DeFi lending protocols. Complementing these findings, \cite{Solmaz2025OptimisticMEV} empirically analyzes speculative MEV opportunities across Ethereum Layer-2 networks from an arbitrage perspective, showing how latency-sensitive arbitrage strategies contribute to persistent demand for blockspace on optimistic rollups. Unlike our work, their study does not investigate liquidation-based strategies or oracle-driven cross-chain liquidation opportunities.

Besides providing empirical evidence and analysis of speculative liquidations in the wild, our work differs from prior research in several important ways. First, while previous studies primarily focus on oracle accuracy, arbitrage, or liquidation competition within a single network, we empirically analyze exploitable timing windows between Chainlink oracle updates across multiple blockchain networks. Second, rather than only studying oracle deviations relative to CEX, we investigate whether oracle updates on one network can be used as predictive signals for upcoming liquidations on another network. Third, we focus specifically on the feasibility of speculative cross-chain liquidations in Aave V3 under volatile market conditions, leveraging both timing asymmetries and transaction inclusion strategies. To the best of our knowledge, this is the first empirical study to demonstrate how liquidators can exploit cross-chain oracle update latency to anticipate and execute profitable liquidations on stale collateralized positions.

\section{Data Collection}
\label{sec:data_collection}

In this section, we describe our data collection process and provide an overview of the collected liquidation events, including the associated on-chain and off-chain price updates observed on October 10, 2025, which, at the time of writing, represents the largest single-day liquidation volume in crypto-asset market history measured in USD.

\subsection{Aave Liquidations}

\begin{figure}[t]
    \centering
    \includegraphics[width=1.0\linewidth]{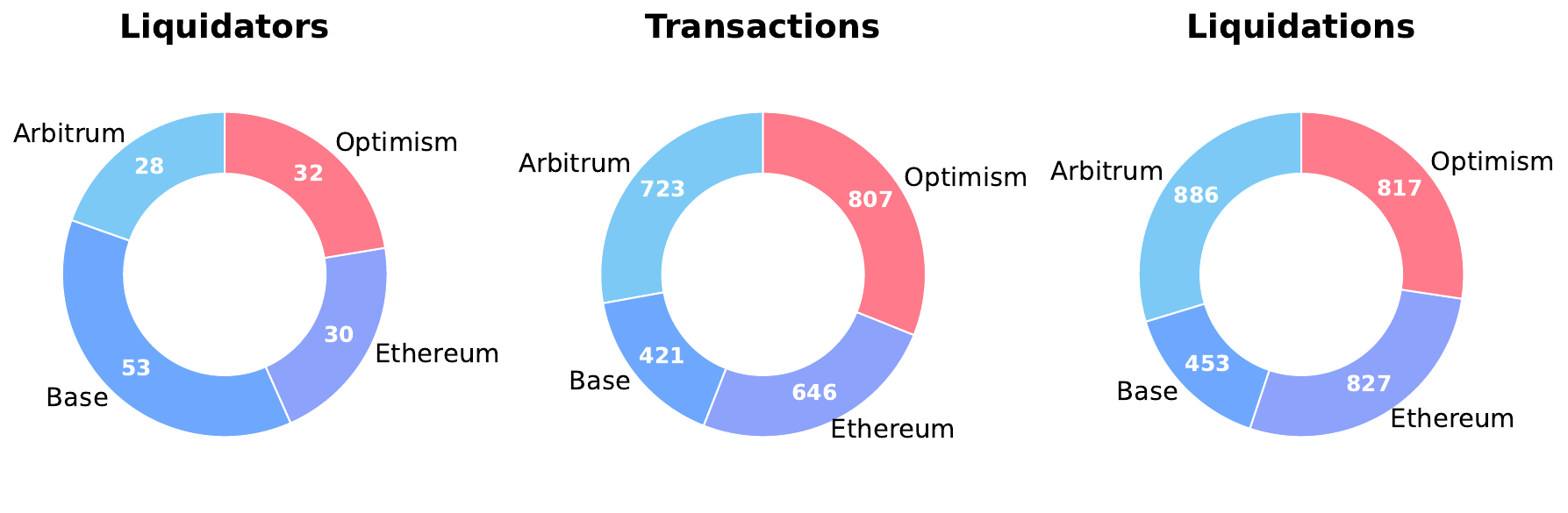}
    \caption{Overview of the number of liquidators, liquidation transactions, and liquidation events detected on October 10, 2025, across Arbitrum, Base, Ethereum, and Optimism.}
    \label{fig:liquidations_overview}
\end{figure}

We collect all Aave liquidations that occurred on October 10, 2025, across Arbitrum, Base, Ethereum, and Optimism using Torres et al.'s open-source scripts \cite{Torres2025}. We retain only liquidation events associated with Aave by filtering for the event topic \texttt{LiquidationCall}. The resulting dataset contains transaction- and block-level information, as well as loan- and liquidation-specific details, including collateral and debt assets, liquidation amounts, and profits realized by liquidators.
\figureautorefname{} \ref{fig:liquidations_overview} summarizes the number of liquidators, liquidation transactions, and liquidation events identified across the analyzed networks. While Base exhibits the largest number of distinct liquidators (53), the highest number of liquidation transactions (807) is observed on Optimism. In contrast, Arbitrum records the largest number of individual liquidation events (886). This suggests that Base experiences the strongest competition among liquidators, Optimism the highest liquidation activity in terms of transactions, and Arbitrum a higher prevalence of liquidators optimizing their transactions to perform multiple liquidations at the same time.

\begin{table}[b]
\centering
\begin{adjustbox}{max width=\columnwidth}
\begin{tabular}{lrrrr}
\toprule
\textbf{Chain} & \textbf{Price Feed Contracts} & \textbf{Total Events} & \textbf{NewTransmission Events} & \textbf{DON Price Observations} \\
\midrule
Arbitrum & 9 & \num{12672} & \num{4224} & \num{42130} \\
Base & 11 & \num{6660} & \num{2220} & \num{22193} \\
Ethereum & 32 & \num{4215} & \num{1405} & \num{27509} \\
Optimism & 11 & \num{12771} & \num{4257} & \num{41052} \\
\bottomrule
\end{tabular}
\end{adjustbox}
\caption{Overview of the collected Chainlink price oracle events and DON price observations across all analyzed chains on October 10, 2025.}
\label{tab:data_overview}
\end{table}

\subsection{Chainlink Price Updates and DON Observations}

We first collected the Chainlink oracle addresses involved in the computation of liquidation HFs for Aave v3 collateralized positions across Ethereum, Arbitrum, Optimism, and Base on October 10, 2025. For each detected liquidation event, we executed a sequence of smart contract calls to recover the underlying oracle infrastructure associated with the liquidation.
Specifically, we first queried the Aave lending pool contract that emitted the liquidation event in order to retrieve the corresponding Aave address provider contract. Afterwards, we queried the address provider to obtain the address of the Aave price oracle. However, this contract does not directly correspond to a Chainlink oracle, but instead serves as an Aave oracle proxy capable of aggregating data from multiple oracle sources.
Using the Aave oracle contract, we then invoked the \texttt{getSourceOfAsset()} function for both the debt asset and the collateral asset associated with the liquidation. This procedure returned the corresponding Chainlink oracle proxy addresses for the respective assets. Finally, by invoking the \texttt{aggregator()} function on each Chainlink proxy contract, we retrieved the underlying Chainlink price feed contract responsible for receiving off-chain price updates via \texttt{NewTransmission} events.

\begin{figure}[b]
    \centering
    \includegraphics[width=1.0\linewidth,height=5cm]{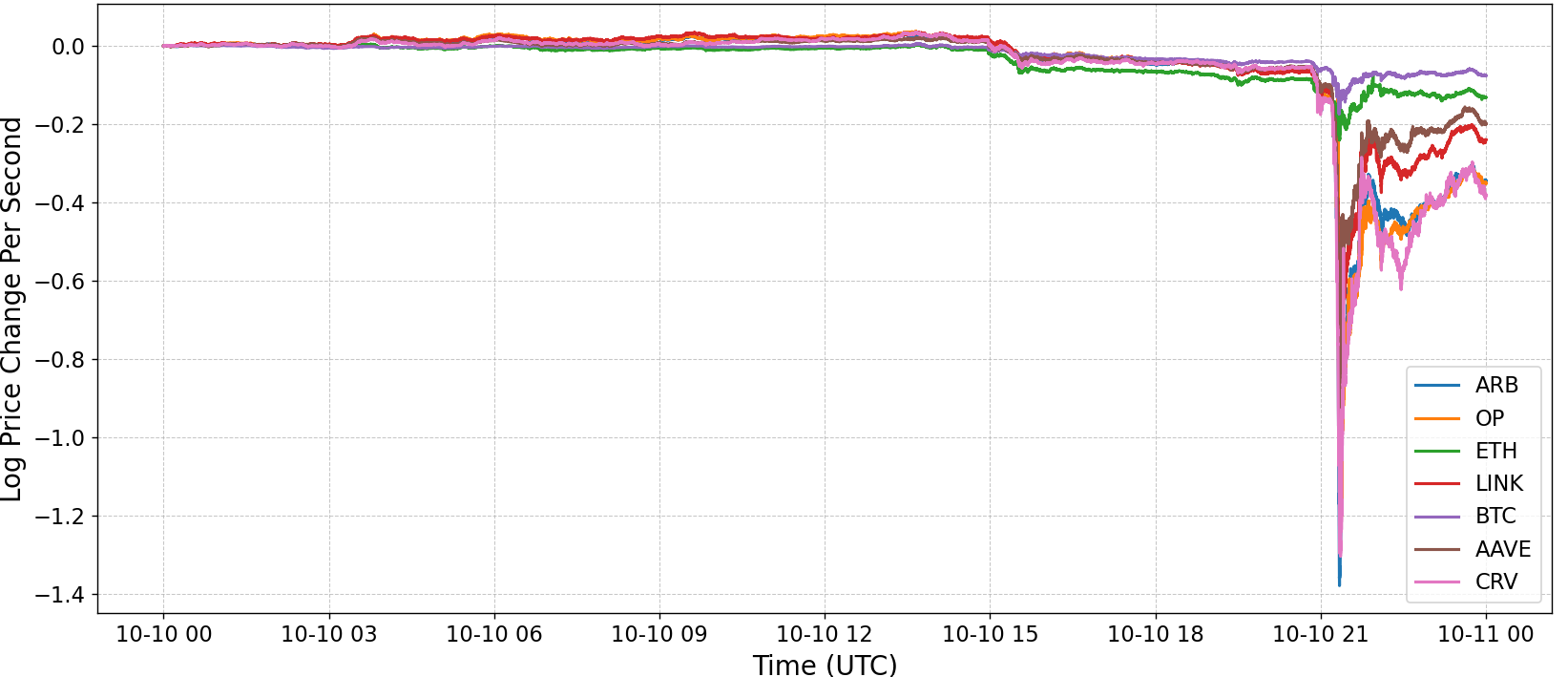}
    \caption{Logarithmic price changes of asset prices for Binance USDT trading pairs on October 10, 2025, sampled at one-second intervals. ETH and BTC exhibited comparatively lower volatility than the remaining analyzed assets for that particular day.}
    \label{fig:binance_price}
\end{figure}

In addition to identifying the price oracle contracts associated with liquidation events, we queried the price feed events (\texttt{NewTransmission}, \texttt{NewRound}, and \texttt{AnswerUpdated}) emitted by the corresponding Chainlink aggregator contracts through our archive nodes using a custom Python script built using the \textit{web3} library.
The collected events include both the aggregated oracle prices of the corresponding assets and the individual price \texttt{observations} submitted by Decentralized Oracle Network (DON) peers. Consequently, our dataset comprises all \texttt{NewTransmission} events associated with the relevant Chainlink aggregator contracts, including raw \texttt{observations} arrays, transmitter identities, and oracle configuration parameters such as deviation thresholds (DT), heartbeat intervals (HB), SVR compatibility, and DON peer lists obtained from on-chain contracts and publicly available Chainlink APIs. An overview of the resulting dataset is provided in Table \ref{tab:data_overview}.

\subsection{Binance Spot Prices}

We additionally collected Binance spot market prices for ARB, OP, ETH, LINK, BTC, AAVE, and CRV on October 10, 2025 using the Binance API. The dataset comprises second-level interval data, including open, close, low, and high prices, as well as trading volume for each of the considered assets. Consistent with prior work, we rely on USDT-denominated trading pairs, as Binance USDT pairs exhibit the highest CEX trading volume among fiat-USD-pegged assets within cryptocurrency markets. The logarithmic price changes of the collected Binance spot prices are illustrated in Figure \ref{fig:binance_price}.

\section{Oracle and Liquidation Landscape}
\label{sec:landscape}

\begin{table}[t]
    \centering
    \begin{adjustbox}{max width=\columnwidth}
    \begin{tabular}{@{}l c r r r r r r r@{}}
        \toprule
        \textbf{Price Feed} & \textbf{SVR} & \textbf{HB\,(s)} & \textbf{DT} & \textbf{Updates} & \textbf{Liquidations} & \textbf{Observations} & \textbf{Peers} & \textbf{Same Obs.} \\
        \midrule
USDC / USD & \cmark & 86{,}400 & 0.25\% & 3 & 300 & 63 & 16 & 39.7\% \\
ETH / USD & \cmark & 3{,}600 & 0.50\% & 122 & 258 & 3{,}765 & 31 & 59.5\% \\
USDT / USD & \xmark & 86{,}400 & 0.25\% & 4 & 213 & 64 & 16 & 26.6\% \\
LINK / USD & \cmark & 3{,}600 & 1.00\% & 114 & 111 & 3{,}493 & 31 & 56.4\% \\
UNI / USD & \xmark & 3{,}600 & 1.00\% & 121 & 87 & 2{,}299 & 19 & 38.5\% \\
BTC / USD & \cmark & 3{,}600 & 0.50\% & 79 & 85 & 2{,}503 & 31 & 55.1\% \\
AAVE / USD & \cmark & 3{,}600 & 1.00\% & 92 & 76 & 1{,}742 & 19 & 46.3\% \\
wBTC / BTC & \xmark & 86{,}400 & 0.50\% & 1 & 64 & 16 & 16 & 68.8\% \\
ETH / USD & \cmark & 3{,}600 & 0.50\% & 117 & 45 & 3{,}606 & 31 & 59.9\% \\
CRV / USD & \xmark & 86{,}400 & 1.00\% & 140 & 38 & 2{,}643 & 19 & 41.4\% \\
DAI / USD & \xmark & 86{,}400 & 0.25\% & 27 & 28 & 511 & 19 & 45.4\% \\
ENS / USD & \xmark & 86{,}400 & 2.00\% & 56 & 13 & 869 & 16 & 38.2\% \\
LDO / ETH & \xmark & 86{,}400 & 2.00\% & 43 & 12 & 688 & 16 & 44.6\% \\
EURC / USD & \xmark & 86{,}400 & 0.30\% & 6 & 12 & 90 & 16 & 68.7\% \\
EUR / USD & \xmark & 86{,}400 & 0.15\% & 3 & 12 & 48 & 16 & 66.7\% \\
BTC / USD & \cmark & 3{,}600 & 0.50\% & 81 & 9 & 2{,}437 & 31 & 54.7\% \\
LINK / ETH & \xmark & 3{,}600 & 1.00\% & 60 & 8 & 920 & 16 & 36.5\% \\
1INCH / USD & \xmark & 86{,}400 & 2.00\% & 85 & 7 & 1{,}552 & 19 & 38.9\% \\
USDtb / USD & \xmark & 86{,}400 & 0.50\% & 1 & 4 & 16 & 16 & 75.0\% \\
USDT / ETH & \xmark & 86{,}400 & 1.00\% & 38 & 2 & 608 & 16 & 41.8\% \\
BAL / USD & \xmark & 86{,}400 & 2.00\% & 30 & 2 & 480 & 16 & 39.4\% \\
AAVE / ETH & \xmark & 86{,}400 & 2.00\% & 25 & 2 & 400 & 16 & 42.5\% \\
BTC / ETH & \xmark & 3{,}600 & 0.50\% & 8 & 2 & 124 & 16 & 35.2\% \\
USDC / USD & \cmark & 86{,}400 & 0.25\% & 4 & 2 & 47 & 16 & 40.6\% \\
PYUSD / USD & \xmark & 86{,}400 & 0.30\% & 3 & 2 & 48 & 16 & 16.7\% \\
USDs / USD & \xmark & 82{,}800 & 0.30\% & 1 & 2 & 16 & 16 & 75.0\% \\
RLUSD / USD & \xmark & 86{,}400 & 0.30\% & 1 & 2 & 16 & 16 & 56.2\% \\
SNX / ETH & \xmark & 86{,}400 & 2.00\% & 67 & 1 & 1{,}064 & 16 & 45.9\% \\
ENJ / ETH & \xmark & 86{,}400 & 2.00\% & 49 & 1 & 757 & 16 & 32.5\% \\
KNC / ETH & \xmark & 86{,}400 & 2.00\% & 22 & 1 & 351 & 16 & 30.2\% \\
LUSD / USD & \xmark & 86{,}400 & 1.00\% & 1 & 1 & 19 & 19 & 42.1\% \\
USDP / USD & \xmark & 86{,}400 & 1.00\% & 1 & 1 & 19 & 19 & 68.4\% \\
        \bottomrule
    \end{tabular}
    \end{adjustbox}
    \caption{Ethereum Chainlink oracle configurations and associated statistics for price feeds involved in Aave~V3 liquidations on October 10, 2025.
    }
    \label{tab:ethereum_liquidation_stats}
\end{table}

\begin{table}
    \centering
    \begin{adjustbox}{max width=\columnwidth}
    \begin{tabular}{@{}l c r r r r r r r@{}}
        \toprule
\textbf{Price Feed} & \textbf{SVR} & \textbf{HB\,(s)} & \textbf{DT} & \textbf{Updates} & \textbf{Liquidations} & \textbf{Observations} & \textbf{Peers} & \textbf{Same Obs.} \\
\midrule
USDC / USD    & \xmark & 86{,}400 & 0.30\% & 4   & 351 & 40     & 10 & 25.0\% \\
ETH / USD     & \xmark & 1{,}200  & 0.15\% & 480 & 249 & 4{,}799 & 10 & 30.2\% \\
BTC / USD     & \xmark & 1{,}200  & 0.10\% & 444 & 91  & 4{,}439 & 10 & 24.0\% \\
AAVE / USD    & \xmark & 86{,}400 & 0.50\% & 185 & 61  & 1{,}850 & 10 & 24.1\% \\
EUR / USD     & \xmark & 3{,}600  & 0.10\% & 31  & 41  & 308    & 10 & 41.8\% \\
EURC / USD    & \xmark & 3{,}600  & 0.10\% & 6   & 41  & 60     & 10 & 60.0\% \\
AERO / USD    & \xmark & 86{,}400 & 0.50\% & 191 & 13  & 1{,}910 & 10 & 33.4\% \\
VIRTUAL / USD & \xmark & 86{,}400 & 0.50\% & 273 & 4   & 2{,}729 & 10 & 25.5\% \\
wstETH / ETH  & \xmark & 86{,}400 & 0.50\% & 1   & 1   & 10     & 10 & 90.0\% \\
BTC / USD     & \xmark & 1{,}200  & 0.10\% & 444 & 1   & 4{,}439 & 10 & 24.0\% \\
cbBTC / USD   & \xmark & 1{,}200  & 0.50\% & 161 & 1   & 1{,}609 & 9  & 64.0\% \\
        \bottomrule
    \end{tabular}
    \end{adjustbox}
    \caption{Base Chainlink oracle configurations and associated statistics for price feeds involved in Aave~V3 liquidations on October 10, 2025.}
    \label{tab:base_liquidation_stats}
\end{table}

\begin{table}
    \centering
    \begin{adjustbox}{max width=\columnwidth}
    \begin{tabular}{@{}l c r r r r r r r@{}}
        \toprule
\textbf{Price Feed} & \textbf{SVR} & \textbf{HB\,(s)} & \textbf{DT} & \textbf{Updates} & \textbf{Liquidations} & \textbf{Observations} & \textbf{Peers} & \textbf{Same Obs.} \\
\midrule
USDC / USD  & \xmark & 86{,}400 & 0.10\% & 14   & 501 & 140    & 10 & 25.0\% \\
ARB / USD   & \xmark & 86{,}400 & 0.05\% & 1,475 & 280 & 14{,}647 & 10 & 29.0\% \\
ETH / USD   & \xmark & 86{,}400 & 0.05\% & 1,193 & 203 & 11{,}925 & 10 & 34.0\% \\
USDT / USD  & \xmark & 86{,}400 & 0.10\% & 28   & 177 & 280    & 10 & 16.4\% \\
LINK / USD  & \xmark & 3{,}600  & 0.02\% & 513  & 129 & 5{,}128  & 10 & 27.3\% \\
BTC / USD   & \xmark & 86{,}400 & 0.05\% & 784  & 103 & 7{,}840  & 10 & 28.8\% \\
AAVE / USD  & \xmark & 86{,}400 & 0.05\% & 201  & 71  & 2{,}010  & 10 & 29.4\% \\
DAI / USD   & \xmark & 86{,}400 & 0.10\% & 14   & 11  & 140    & 10 & 25.0\% \\
LUSD / USD  & \xmark & 86{,}400 & 0.30\% & 2    & 1   & 20     & 10 & 15.0\% \\
        \bottomrule
    \end{tabular}
    \end{adjustbox}
    \caption{Arbitrum Chainlink oracle configurations and associated statistics for price feeds involved in Aave~V3 liquidations on October 10, 2025.}
    \label{tab:arbitrum_liquidation_stats}
\end{table}

\begin{table}
    \centering
    \begin{adjustbox}{max width=\columnwidth}
    \begin{tabular}{@{}l c r r r r r r r@{}}
        \toprule
\textbf{Price Feed} & \textbf{SVR} & \textbf{HB\,(s)} & \textbf{DT} & \textbf{Updates} & \textbf{Liquidations} & \textbf{Observations} & \textbf{Peers} & \textbf{Same Obs.} \\
\midrule
OP / USD     & \xmark & 1{,}200  & 0.02\% & 689 & 475 & 6{,}877 & 10 & 23.7\% \\
USDC / USD   & \xmark & 86{,}400 & 0.01\% & 9   & 479 & 90     & 10 & 27.8\% \\
ETH / USD    & \xmark & 1{,}200  & 0.15\% & 611 & 216 & 6{,}102 & 10 & 29.4\% \\
USDT / USD   & \xmark & 86{,}400 & 0.10\% & 16  & 191 & 160    & 10 & 26.2\% \\
LINK / USD   & \xmark & 1{,}200  & 0.20\% & 609 & 60  & 5{,}592 & 10 & 26.3\% \\
AAVE / USD   & \xmark & 1{,}200  & 0.20\% & 595 & 55  & 5{,}443 & 10 & 29.0\% \\
BTC / USD    & \xmark & 1{,}200  & 0.10\% & 530 & 38  & 4{,}847 & 10 & 29.8\% \\
DAI / USD    & \xmark & 86{,}400 & 0.10\% & 11  & 38  & 110    & 10 & 18.2\% \\
SNX / USD    & \xmark & 1{,}200  & 0.10\% & 938 & 19  & 9{,}347 & 10 & 24.3\% \\
SUSD / USD   & \xmark & 1{,}200  & 0.10\% & 2   & 4   & 19     & 10 & 57.8\% \\
WBTC / USD   & \xmark & 1{,}200  & 0.10\% & 247 & 1   & 2{,}465 & 10 & 47.3\% \\
        \bottomrule
    \end{tabular}
    \end{adjustbox}
    \caption{Optimism Chainlink oracle configurations and associated statistics for price feeds involved in Aave~V3 liquidations on October 10, 2025.}
    \label{tab:optimism_liquidation_stats}
\end{table}

\Cref{tab:ethereum_liquidation_stats}, \Cref{tab:base_liquidation_stats}, \Cref{tab:arbitrum_liquidation_stats}, and \Cref{tab:optimism_liquidation_stats} summarize oracle configuration parameters and aggregate liquidation statistics for all price feeds linked to the identified liquidations across all four chains. We observe substantial variation in heartbeat (HB) intervals and deviation thresholds (DT) across chains, with Layer-2 networks generally exhibiting more frequent oracle updates due to their lower block times and reduced transaction costs. 

The tables further provide a cross-chain overview of submitted observations and liquidation counts. Price feeds are ranked according to the number of liquidations.
In addition, the tables report the number of peers participating in each DON and compare the proportion of identical versus unique observations submitted by DON participants. Two observations are considered identical if and only if they report the exact same raw price integer, with no tolerance for rounding differences. Hence, 
a high ratio of same observations suggests that multiple oracle peers rely on highly overlapping or identical off-chain price sources. Feeds sharing the same name on the same chain correspond to distinct oracle contracts (different proxy addresses).

As expected, we observe that stablecoin price feeds, such as USDC/USD and USDT/USD, exhibit substantially fewer oracle updates and observations than more volatile crypto-asset feeds such as ETH/USD, ARB/USD, and OP/USD, due to the price of stablecoins being most of the time stable. In particular, lower market-cap and more volatile assets generate the highest number of oracle updates, as exemplified by the ARB/USD feed with 14{,}647 observations during the study period. This behavior is primarily driven by the deviation threshold mechanism, under which increased price volatility causes oracle updates to be triggered more frequently. However, we also observe substantial variation across price feeds in the proportion of DON peers submitting identical price observations. For example, the lowest percentage of identical observations is observed for the LUSD/USD price feed on Arbitrum, where only 15\% of submitted observations are identical, whereas the highest percentage is observed for the wstETH/ETH price feed on Base, where approximately 90\% of observations share identical values. This suggests that, for certain feeds, DON peers may rely on highly overlapping or insufficiently diverse off-chain data sources.

In contrast, liquidation counts are distributed more evenly across both stable and volatile price feeds. This is because the computation of Aave~V3 HFs depends simultaneously on both collateral and debt asset prices. For example, a position collateralized with USDT while borrowing ETH depends on both the USDT/USD and ETH/USD oracle feeds. Consequently, liquidations involving a given position contribute to the liquidation statistics of multiple related price feeds, leading to similar liquidation counts across otherwise differently behaving oracle feeds.

\section{Detecting Speculative Liquidators in the Wild}

In this section, we investigate the prevalence of speculative liquidations in the wild by first analyzing the occurrence of liquidations executed within the same block as oracle price updates, and subsequently examining the prevalence and behavioral characteristics of speculative and non-speculative liquidators.

\subsection{Analysis of Same-Block Liquidations}

Rollup chains such as Arbitrum, Base, and Optimism do not expose a public mempool. As a result, users cannot observe pending transactions, including pending price oracle update transactions, and therefore cannot directly backrun oracle updates within the same block. Nevertheless, speculative liquidators seeking to extract OEV may attempt to anticipate pending oracle updates and continuously submit liquidation transactions, aiming to have their transactions included in the same block as the corresponding price update.

To investigate this phenomenon, we analyze the occurrence of same-block liquidations, i.e., liquidation transactions that are included in the same block as their associated Chainlink oracle price update transaction. Our methodology leverages the Chainlink price oracle contract addresses associated with individual liquidation transactions identified during the data collection process (see \sectionautorefname{} \ref{sec:data_collection}) in order to detect same-block liquidations.
 Using our dataset of collected \texttt{NewTransmission} events, i.e., Chainlink price oracle updates, we first group both oracle update transactions and liquidation transactions by block number.
Next, for each liquidation transaction, we discard all oracle update transactions within the same block that occur after the liquidation transaction. This filtering is performed using the transaction index, as an oracle update that is executed after a liquidation cannot have influenced that liquidation event.
Finally, we determine whether any of the oracle contract addresses associated with the liquidation match the oracle contract address of a preceding price update transaction within the same block. A match indicates that the liquidation was triggered by that oracle update.

\begin{table}[t]
    \centering
    \begin{adjustbox}{max width=\columnwidth}
    \begin{tabular}{l r r r r r r r}
        \toprule
        & & & \multicolumn{5}{c}{\textbf{Transaction Distance}} \\
        \cline{4-8}
        \textbf{Chain} & \textbf{Liquidations} & \textbf{Same-Block} & \textbf{Bundled ($d_{tx} = 1$)} & \textbf{Min} & \textbf{Mean} & \textbf{Median} & \textbf{Max}\\
        \midrule
        Arbitrum & 886 & 169 (19.08\%) & 36 (53.73\%) & 1 & 2.52 & 1.00 & 14 \\
        Base     & 453 & 169 (37.31\%) & 8 (5.56\%) & 1 & 109.22 & 29.50 & 1,697 \\
        Optimism & 817 &  80 (9.79\%) & 7 (9.21\%) & 1 & 55.40 & 37.00 & 306 \\
        \bottomrule
    \end{tabular}
    \end{adjustbox}
    \caption{Percentage of Aave liquidations executed within the same block as their corresponding Chainlink price oracle updates, along with statistics on transaction distance (i.e., the positional difference between the oracle update transaction and the liquidation transaction) across chains employing no public mempools. 
    }
    \label{tab:liq_block_match}
\end{table}

\Cref{tab:liq_block_match} reports the total number of liquidations identified on October 10th, 2025 across Arbitrum, Base, and Optimism, together with the proportion of liquidations that occurred within the same block as their associated price oracle update transaction. The table additionally presents statistics on the transaction distance between the oracle update transaction and the liquidation transaction for these same-block liquidations. A transaction distance of one indicates that the liquidation transaction immediately follows the oracle update transaction within the block.

Among the analyzed chains, Base exhibits the highest proportion of same-block liquidations at 37.31\%. However, the median transaction distance on Base is 29.50, suggesting the presence of speculative OEV behavior in which liquidators submit transactions in anticipation of pending oracle updates. In contrast, Arbitrum shows a lower proportion of same-block liquidations at 19.08\%, but 53.73\% of these liquidations are back-to-back with the oracle update transaction (see the ``Bundled'' column), corresponding to a transaction distance of one. This behavior is further reflected in Arbitrum’s median transaction distance of one. However, it is important to note that Arbitrum employs a substantially shorter block time of approximately 250 ms, whereas Base and Optimism operate with block times of roughly 2 seconds. Consequently, Arbitrum blocks are typically smaller and contain fewer transactions, which may explain the considerably lower observed transaction distances relative to Base and Optimism and the ease for a liquidation transaction to be directly included after Chainlink price oracle update when emitting multiple liquidation transactions at the same time.
 Optimism exhibits the lowest proportion of same-block liquidations at 9.67\%. 
Overall, the presence of a non-negligible number of same-block liquidations across all three chains, despite the absence of a public mempool, suggests that speculative OEV extraction may already be occurring in practice. 

\begin{table}[b]
    \centering
    \begin{adjustbox}{max width=\columnwidth}
    \begin{tabular}{l r r | r r | r r}
        \toprule
         & \multicolumn{2}{c}{\textbf{Liquidators}} & \multicolumn{2}{c}{\textbf{Transactions}} & \multicolumn{2}{c}{\textbf{Liquidations}} \\
        \cline{2-7}
        \textbf{Chain} & \textbf{Speculative} & \textbf{Non-Spec.} & \textbf{Speculative} & \textbf{Non-Spec.} & \textbf{Speculative} & \textbf{Non-Spec.} \\
        \midrule
        Arbitrum & 12 (42.86\%) & 16 (57.14\%) & 291 (40.25\%) & 432 (59.75\%) & 447 (50.45\%) & 439 (49.55\%) \\
        Base &  34 (64.15\%) & 19 (35.85\%) & 161 (38.24\%) & 260 (61.76\%) & 193 (42.60\%) & 260 (57.40\%) \\
        Optimism & 18 (56.25\%) & 14 (43.75\%) & 186 (23.05\%) & 621 (76.95\%) & 191 (23.38\%) & 626 (76.62\%) \\
        \midrule
        \textbf{Total} & \textbf{64 (56.64\%)} & \textbf{49 (43.36\%)} & \textbf{638 (32.70\%)} & \textbf{1,313 (67.30\%)} & \textbf{831 (38.54\%)} & \textbf{1,325 (61.46\%)} \\
        \bottomrule
    \end{tabular}
    \end{adjustbox}
    \caption{Overview of identified speculative and non-speculative liquidators, liquidation transactions, and liquidation events across each chain.}
    \label{tab:speculative_liquidation_overview}
\end{table}

\subsection{Detection of Speculative Liquidations and Liquidators}

Given the large number of detected same-block liquidations, we next seek to empirically measure and identify evidence of speculative liquidation behavior.
For each of the 2,986 identified liquidation transactions, we collect all transactions contained within the preceding 10 blocks as well as the subsequent 10 blocks surrounding the liquidation event. From this set, we retain only transactions that share both the same destination address and identical input calldata as the successful liquidation transaction. Subsequently, a liquidation transaction is classified as speculative if at least one of the following heuristics is satisfied: 
(i) The block containing the successful liquidation transaction also contains at least one additional transaction with the same destination address and identical input calldata;
(ii) at least one directly adjacent block (i.e., with a block distance of one preceding or succeeding the liquidation block) contains at least one transaction with the same destination address and identical input calldata as the successful liquidation transaction.
A transaction destination address (i.e., a liquidator) is classified as speculative if at least one of its successful liquidation transactions is identified as speculative according to the two heuristics above, and all successful liquidations associated with that destination address are likewise classified as speculative.

\begin{table}
    \centering
    \begin{adjustbox}{max width=\columnwidth}
    \begin{tabular}{@{}lcrrrrr@{}}
        \toprule
        \textbf{Chain} & \textbf{Liquidator}
            & \textbf{Liquidations}
            & \textbf{Same-Block} & \textbf{Oracle Distance} \\
        \midrule
        \multirow{10}{*}{Arbitrum}
            & \cellcolor{gray!20}\href{https://arbiscan.io/address/0xb9516c655831917704A5000bA6f4EB64E816DC2d}{\texttt{0xb9516c655831917704A5000bA6f4EB64E816DC2d}} & 
            \cellcolor{gray!20}
            215 (24.27\%) & 
            \cellcolor{gray!20}
            110 (12.42\%) &
            \cellcolor{gray!20} 1.00
            \\
            & \href{https://arbiscan.io/address/0x001f8151FC6d0a14608B48F1d2a2AeA66c80cF38}{\texttt{0x001f8151FC6d0a14608B48F1d2a2AeA66c80cF38}} & 
            141 (15.91\%) & 
            0 (0.00\%) & - \\
            & \href{https://arbiscan.io/address/0x48daab9f7Ed6E3184c26D9dDACcc356D52D6237F}{\texttt{0x48daab9f7Ed6E3184c26D9dDACcc356D52D6237F}} & 
            95 (10.72\%) & 1 (0.11\%) & 2.00 \\
            & \cellcolor{gray!20}
            \href{https://arbiscan.io/address/0xd249aB6aaC55a7AD0ceF7fCFD672d47387D7e70F}{\texttt{0xd249aB6aaC55a7AD0ceF7fCFD672d47387D7e70F}} & 
            \cellcolor{gray!20}
            87 (9.82\%) & \cellcolor{gray!20}
            21 (2.37\%) & \cellcolor{gray!20} 1.00
            \\
            & \cellcolor{gray!20}
            \href{https://arbiscan.io/address/0x6290c280f1393d33f04d6A59993cbe7d3ECCDFcF}{\texttt{0x6290c280f1393d33f04d6A59993cbe7d3ECCDFcF}} & 
            \cellcolor{gray!20}
            79 (8.92\%) & \cellcolor{gray!20}
            25 (2.82\%) & \cellcolor{gray!20} 3.00
            \\
            & \href{https://arbiscan.io/address/0x545875b6975eb527726C87B82B5f031ED861De8A}{\texttt{0x545875b6975eb527726C87B82B5f031ED861De8A}} & 
            59 (6.66\%) & 0 (0.00\%) & - \\
            & \cellcolor{gray!20}
            \href{https://arbiscan.io/address/0x9f83664976aDc0abe9C375151B0E5B36eE7703D7}{\texttt{0x9f83664976aDc0abe9C375151B0E5B36eE7703D7}} & 
            \cellcolor{gray!20}
            52 (5.87\%) & \cellcolor{gray!20}
            0 (0.00\%) & \cellcolor{gray!20} -
             \\
            & \href{https://arbiscan.io/address/0xa441ed20E5A5c51C067549316B86F84CC988D284}{\texttt{0xa441ed20E5A5c51C067549316B86F84CC988D284}} & 
            27 (3.05\%) & 0 (0.00\%) & - \\
            & \href{https://arbiscan.io/address/0x4044E9E4E9d0ef73026106d097DCdB1d609436A7}{\texttt{0x4044E9E4E9d0ef73026106d097DCdB1d609436A7}} & 
            22 (2.48\%) & 0 (0.00\%) & - \\
            & \href{https://arbiscan.io/address/0x4Ae48f73Abc9F6f7A918ff405584B9314A5df0b8}{\texttt{0x4Ae48f73Abc9F6f7A918ff405584B9314A5df0b8}} & 
            19 (2.15\%) & 0 (0.00\%) & - \\
        \midrule
        
        \multirow{10}{*}{Base}
        & \href{https://basescan.org/address/0xD251c1325c5d7b29C6219912D8648a3149cDF57B}{\texttt{0xD251c1325c5d7b29C6219912D8648a3149cDF57B}} & 
        100 (22.08\%) & 0 (0.00\%) & - \\
        & \cellcolor{gray!20}\href{https://basescan.org/address/0xc89c328609aB58E256Cd2b5aB4F4aF2EFb9fcA33}{\texttt{0xc89c328609aB58E256Cd2b5aB4F4aF2EFb9fcA33}} & 
        \cellcolor{gray!20}53 (11.70\%) & \cellcolor{gray!20} 23 (5.08\%) & \cellcolor{gray!20} 139.00 \\
        & \cellcolor{gray!20}\href{https://basescan.org/address/0x3ba19cB44eF72B0B198325E17623BcE10Bae2753}{\texttt{0x3ba19cB44eF72B0B198325E17623BcE10Bae2753}} & 
        \cellcolor{gray!20}38 (8.39\%) & \cellcolor{gray!20} 20 (4.42\%) & \cellcolor{gray!20} 28.00 \\
        & \href{https://basescan.org/address/0x964AeE3e4E3BBc7245B33dA097030e95EE408170}{\texttt{0x964AeE3e4E3BBc7245B33dA097030e95EE408170}} & 
        33 (7.29\%) & 20 (4.42\%) & 26.00 \\
        & \cellcolor{gray!20}\href{https://basescan.org/address/0xec5aCd2dfdf3B2e4dcb955C7eC8B2b74605d8E9c}{\texttt{0xec5aCd2dfdf3B2e4dcb955C7eC8B2b74605d8E9c}} & 
        \cellcolor{gray!20}27 (5.96\%) & \cellcolor{gray!20} 23 (5.08\%) & \cellcolor{gray!20} 2.00 \\
        & \cellcolor{gray!20}\href{https://basescan.org/address/0x888888887A487f209e31a692B227d8D1ff9070ba}{\texttt{0x888888887A487f209e31a692B227d8D1ff9070ba}} & 
        \cellcolor{gray!20}22 (4.86\%) & \cellcolor{gray!20} 1 (0.22\%) & \cellcolor{gray!20} 149.00 \\
        & \href{https://basescan.org/address/0x3347277366deCC91d65D2762E792E6BaA471b805}{\texttt{0x3347277366deCC91d65D2762E792E6BaA471b805}} & 
        21 (4.64\%) & 8 (1.77\%) & 26.00 \\
        & \href{https://basescan.org/address/0xaDDD8ECec572E3BA9975578e1927221eC25Dab50}{\texttt{0xaDDD8ECec572E3BA9975578e1927221eC25Dab50}} & 
        18 (3.97\%) & 0 (0.00\%) & - \\
        & \href{https://basescan.org/address/0x3f482cA03E6d5FA63884918d5dC379F14cA0aD86}{\texttt{0x3f482cA03E6d5FA63884918d5dC379F14cA0aD86}} & 
        17 (3.75\%) & 14 (3.09\%) & 17.50 \\
        & \href{https://basescan.org/address/0xf4Cd829526070C2C0af9E6Dce2C9255130f7ec68}{\texttt{0xf4Cd829526070C2C0af9E6Dce2C9255130f7ec68}} & 
        14 (3.09\%) & 0 (0.00\%) & - \\
        \midrule
        
        \multirow{10}{*}{Optimism} & 
        \href{https://optimistic.etherscan.io/address/0x7518Ba8b4021D83caFB1e46Fa1250f54fA86b6Ab}{\texttt{0x7518Ba8b4021D83caFB1e46Fa1250f54fA86b6Ab}} & 
        170 (27.16\%) & 8 (1.28\%) & 70.00 \\
        & \cellcolor{gray!20}\href{https://optimistic.etherscan.io/address/0x8888888885FcA4862619cAFeda9b03768C930012}{\texttt{0x8888888885FcA4862619cAFeda9b03768C930012}} & 
        \cellcolor{gray!20}112 (17.89\%) & \cellcolor{gray!20} 6 (0.96\%) & \cellcolor{gray!20} 45.00 \\
        & \href{https://optimistic.etherscan.io/address/0x4661f836F07c78a01D5490bFf6e8587B9291C6B8}{\texttt{0x4661f836F07c78a01D5490bFf6e8587B9291C6B8}} & 
        93 (14.86\%) & 5 (0.80\%) & 32.00 \\
        & \href{https://optimistic.etherscan.io/address/0x4db5FBb9246F66bd83661D680F308cc05c5FC3Bf}{\texttt{0x4db5FBb9246F66bd83661D680F308cc05c5FC3Bf}} & 
        88 (14,06\%) & 3 (0.48\%) & 13.00 \\
        & \href{https://optimistic.etherscan.io/address/0x874Bd329Adf526A61FeE5766B9E96c9519aeBb1b}{\texttt{0x874Bd329Adf526A61FeE5766B9E96c9519aeBb1b}} & 
        77 (12,30\%) & 7 (1.12\%) & 43.00 \\
        & \href{https://optimistic.etherscan.io/address/0x1dC68046E71429F646Dd6B2d0c02385aE61c2129}{\texttt{0x1dC68046E71429F646Dd6B2d0c02385aE61c2129}} & 
        53 (8.46\%) & 2 (0.32\%) & 55.50 \\
        & \cellcolor{gray!20}\href{https://optimistic.etherscan.io/address/0x083CfA7FD187Be983ce5D519fE7ae78357779998}{\texttt{0x083CfA7FD187Be983ce5D519fE7ae78357779998}} & 
        \cellcolor{gray!20} 42 (6.71\%) & \cellcolor{gray!20} 4 (0.64\%) & \cellcolor{gray!20} 63.50 \\
        & \href{https://optimistic.etherscan.io/address/0x59c5f1B4FF4fdf04729787EeE84DdBDf787A3033}{\texttt{0x59c5f1B4FF4fdf04729787EeE84DdBDf787A3033}} & 
        37 (5.91\%) & 1 (0.16\%) & 4.00 \\
        & \href{https://optimistic.etherscan.io/address/0x4B1142124c661B48aD7F98d8C7C8db03016B1F88}{\texttt{0x4B1142124c661B48aD7F98d8C7C8db03016B1F88}} & 
        26 (4.15\%) & 1 (0.16\%) & 34.00 \\
        & \href{https://optimistic.etherscan.io/address/0xe3BadF4DCf5908472468e27CF6b438f3F11D9c5D}{\texttt{0xe3BadF4DCf5908472468e27CF6b438f3F11D9c5D}} & 
        26 (4.15\%) & 5 (0.80\%) & 43.00 \\
        \bottomrule
    \end{tabular}
    \end{adjustbox}
    \caption{Liquidation counts, same-block liquidations, and median transaction distance between liquidation transaction and oracle price update transaction for the top 10 liquidators on each chain. Rows highlighted in gray correspond to liquidators exhibiting speculative behavior.}
    \label{tab:top_liquidators} 
\end{table}

\subparagraph*{Speculative vs. Non-Speculative.}
\tableautorefname{} \ref{tab:speculative_liquidation_overview} summarizes the results obtained by applying our methodology to detect speculative liquidations on October 10, 2025. Our analysis reveals that approximately 57\% of all liquidators active on that day employed speculative liquidation strategies. Such behavior is observed across all chains, with the highest concentration occurring on Base, where 64\% of detected liquidators are classified as speculative.

Despite the substantial prevalence of speculative liquidators, non-speculative liquidators appear to achieve greater success in terms of completed liquidation activity. Overall, only 33\% of successful liquidation transactions are classified as speculative, implying that the majority originates from non-speculative actors. A similar trend is reflected in the number of liquidation events, of which only 39\% are speculative. This pattern is particularly pronounced on Optimism, where merely 23\% of successful liquidation transactions are speculative, while the remaining 77\% are attributable to non-speculative liquidators.
Interestingly, the number of speculative liquidation events exceeds the number of speculative liquidation transactions, suggesting that majority of speculative liquidators execute multiple liquidations within a single transaction. Arbitrum provides a representative example of this phenomenon: although 60\% of successful liquidation transactions are classified as non-speculative, the majority of successful liquidation events (51\%) are speculative. This indicates that speculative liquidators on Arbitrum frequently bundle multiple liquidations into individual transactions, thereby increasing the number of liquidation events attributable to speculative strategies.

\subparagraph*{Top Liquidators.}
\tableautorefname{} \ref{tab:top_liquidators} provides an overview of the top 10 liquidators identified on October 10, 2025, ranked by the number of successful liquidations performed. Rows highlighted in gray correspond to liquidators classified by our methodology as speculative.
The most active liquidator on Arbitrum is identified as speculative, having executed 215 successful liquidations. More than half of these liquidations (110) occurred within the same block as the corresponding oracle price update transaction. Furthermore, the median transaction distance between the oracle update and the liquidation transaction is equal to one, indicating that the liquidation transaction directly followed the oracle price update transaction.
In contrast, the top-performing liquidators on Base and Optimism do not appear to employ speculative strategies. Nevertheless, on both chains, the second most successful liquidator by liquidation count is classified as speculative. Similar to the observations on Arbitrum, speculative liquidators on Base and Optimism exhibit substantially higher proportions of liquidations occurring within the same block as the oracle price update transaction. Given that these Layer-2 chains do not expose a public mempool, precisely positioning a liquidation transaction within the same block as the oracle update is challenging without the use of speculative OEV strategies.
Interestingly, the median transaction distance between liquidation transactions and oracle price updates is considerably larger on Base and Optimism than on Arbitrum. This observation is consistent with the differing block production characteristics of these chains. Specifically, Base and Optimism operate with larger blocks that contain substantially more transactions and therefore greater transaction competition, whereas Arbitrum blocks are typically smaller and include fewer transactions.

\begin{table}[]
    \centering
    \begin{adjustbox}{max width=\columnwidth}
    \begin{tabular}{c r r r r r | r r }
        \toprule
        & & \multicolumn{4}{c}{\textbf{Blocks}} & \multicolumn{2}{c}{\textbf{Transactions}} \\
        \cline{3-5} \cline{6-8}
        \textbf{Chain} & \multicolumn{1}{c}{\textbf{Liquidator}} & \textbf{Min} & \textbf{Mean} & \textbf{Median} & \textbf{Max} & \textbf{Per Block} & \textbf{Reverted}  \\
        \midrule
        \multirow{4}{*}{Arbitrum} & \href{https://arbiscan.io/address/0xb9516c655831917704A5000bA6f4EB64E816DC2d}{\texttt{0xb9516c655831917704A5000bA6f4EB64E816DC2d}} &  0 	& 8.38 &	 9.00 	& 15 	& 10.00 & 14,197 (99.05\%) \\
        & \href{https://arbiscan.io/address/0xd249aB6aaC55a7AD0ceF7fCFD672d47387D7e70F}{\texttt{0xd249aB6aaC55a7AD0ceF7fCFD672d47387D7e70F}} & 0 &	 8.44 &	 10.00 	& 18 	& 15.00 & 5943 (96.97\%)\\
        & \href{https://arbiscan.io/address/0x6290c280f1393d33f04d6A59993cbe7d3ECCDFcF}{\texttt{0x6290c280f1393d33f04d6A59993cbe7d3ECCDFcF}} &  2 	& 6.35 &	 4.00 	& 15 &	 7.50 & 1766 (95.72\%)\\
        & \href{https://arbiscan.io/address/0x9f83664976aDc0abe9C375151B0E5B36eE7703D7}{\texttt{0x9f83664976aDc0abe9C375151B0E5B36eE7703D7}} & 2 	& 2.64 &	 3.00 	& 3 	& 1.00 & 24 (31.58\%)\\
        \midrule

        \multirow{4}{*}{Base} & \href{https://basescan.org/address/0xc89c328609aB58E256Cd2b5aB4F4aF2EFb9fcA33}{\texttt{0xc89c328609aB58E256Cd2b5aB4F4aF2EFb9fcA33}} &  0 	& 0.00 &	 0.00 &	 0 	& 2.00 & 0 (0.00\%) \\
        & \href{https://basescan.org/address/0x3ba19cB44eF72B0B198325E17623BcE10Bae2753}{\texttt{0x3ba19cB44eF72B0B198325E17623BcE10Bae2753}} & 2 &	 6.24 	& 5.00 &	 16 &	 3.00 & 415 (89.25\%)  \\
        & \href{https://basescan.org/address/0xec5aCd2dfdf3B2e4dcb955C7eC8B2b74605d8E9c}{\texttt{0xec5aCd2dfdf3B2e4dcb955C7eC8B2b74605d8E9c}} & 11 	& 14.75 &	 12.50 	& 21 	& 382.00 & 0 (0.00\%) \\
        & \href{https://basescan.org/address/0x888888887A487f209e31a692B227d8D1ff9070ba}{\texttt{0x888888887A487f209e31a692B227d8D1ff9070ba}} & 2 	& 2.00 &	 2.00 	& 2 	& 1.00 & 0 (0.00\%) \\
        \midrule
        
        \multirow{2}{*}{Optimism} & \href{https://optimistic.etherscan.io/address/0x8888888885FcA4862619cAFeda9b03768C930012}{\texttt{0x8888888885FcA4862619cAFeda9b03768C930012}} & 0 &	 0.67 & 0.00 &	 2 &	 1.00 & 2 (1.34\%) \\
        & \href{https://optimistic.etherscan.io/address/0x083CfA7FD187Be983ce5D519fE7ae78357779998}{\texttt{0x083CfA7FD187Be983ce5D519fE7ae78357779998}} & 0 	& 0.00 & 	 0.00 &	 0 	& 1.00 & 5 (10.64\%) \\
        \bottomrule
    \end{tabular}
    \end{adjustbox}
    \caption{Overview of consecutive blocks containing speculative transactions, together with the median number of speculative transactions emitted per block and the proportion of reverted transactions among all speculative transactions for the top speculative liquidators on each chain.}
    \label{tab:liquidator_reverted}
\end{table}

\subparagraph*{Liquidator Profits and Costs.}
Overall, we observe that all analyzed liquidators, regardless of whether they employ speculative or non-speculative strategies, remain profitable over the examined period, with none incurring an aggregate net loss (see \tableautorefname{} \ref{tab:liquidator_profits} and \ref{tab:liquidator_costs} in \appendixautorefname{} \ref{sec:appendix_liquidator_profits_costs}). Furthermore, the number of executed liquidations does not necessarily correlate with the highest realized profit. For instance, on Arbitrum, the most profitable liquidator is classified as non-speculative and ranks only sixth in terms of liquidation count, yet achieved a net profit of nearly 1 million USD within a 24-hour period.
More broadly, non-speculative liquidators appear to be more profitable than speculative liquidators. This trend is particularly pronounced on Arbitrum and Optimism, where non-speculative liquidators achieve the highest average profits per liquidation. As expected, speculative liquidators incur the largest total and average transaction costs, especially on Arbitrum and Base. This effect is further reflected in the minimum profit per liquidation, where speculative liquidators may incur losses of several thousand USD on individual liquidations due to extensive transaction spamming and heightened competition among searchers.
In general, liquidators with the highest liquidation activity tend to interact with a considerably more diverse set of price feeds (see \tableautorefname{} \ref{tab:feeds_per_liquidator} in \appendixautorefname{} \ref{sec:appendix_liquidator_price_feeds}). Nevertheless, broader coverage across price feeds does not necessarily translate into higher cumulative profitability.

\subparagraph*{Speculative Strategies.}
An overview of the block- and transaction-level strategies employed by speculative liquidators identified among the top 10 liquidators across Arbitrum, Base, and Optimism is presented in \tableautorefname{} \ref{tab:liquidator_reverted}.
Our analysis shows that the majority of speculative liquidators emit transactions across multiple consecutive blocks. The largest number of consecutive blocks was observed on Base, where a speculative liquidator continuously emitted transactions across 21 consecutive blocks. This value corresponds to the full size of our observation window, consisting of the 10 blocks preceding the liquidation transaction, the block containing the liquidation transaction itself, and the 10 succeeding blocks. Consequently, this observation suggests that the liquidator may have continued emitting identical transactions for an even longer duration beyond the observed interval. The same liquidator also exhibits the highest median number of speculative transactions emitted per block, namely 382 transactions. Interestingly, this liquidator does not employ a transaction reversion strategy, as none of its speculative transactions reverted.
More broadly, speculative liquidators on Arbitrum appear to rely heavily on transaction reversion strategies. In particular, the speculative transactions emitted by the top three speculative liquidators on Arbitrum revert in at least 95\% of cases. Importantly, regardless of whether speculative transactions revert or not, speculative liquidators generate a substantial volume of transactions that do not modify blockchain state. Instead, these transactions merely read state information to identify potential liquidation opportunities and terminate without executing state-changing operations when no profitable opportunity exists. Nevertheless, such transactions are still executed and included on-chain, thereby consuming block space and contributing to increased transaction fees.

\section{Aligning Off-Chain CEX Prices with On-Chain Oracle Observations}
\label{subsec:methods}

In this section, we study the latency between off-chain asset prices observed on CEXes, such as Binance, and their corresponding on-chain Chainlink oracle observations. For each price recorded in a Chainlink \texttt{NewTransmission} event, we align it with Binance spot prices within a 120-second lookback window, discretized into one-second bins (e.g., $[0,1)$). Each observation is uniquely assigned to the bin whose midpoint price exactly matches the observed oracle price.
This matching procedure prevents artificial inflation of counts that would arise if a single oracle observation were mapped to multiple overlapping CEX price intervals. 
While this strict criterion excludes unmatched observations, we prioritize precision over coverage to avoid introducing ambiguity into the alignment process.

\begin{figure}
    \centering
    \begin{subfigure}[b]{1.0\textwidth}
        \includegraphics[width=1.0\linewidth,height=5.5cm]{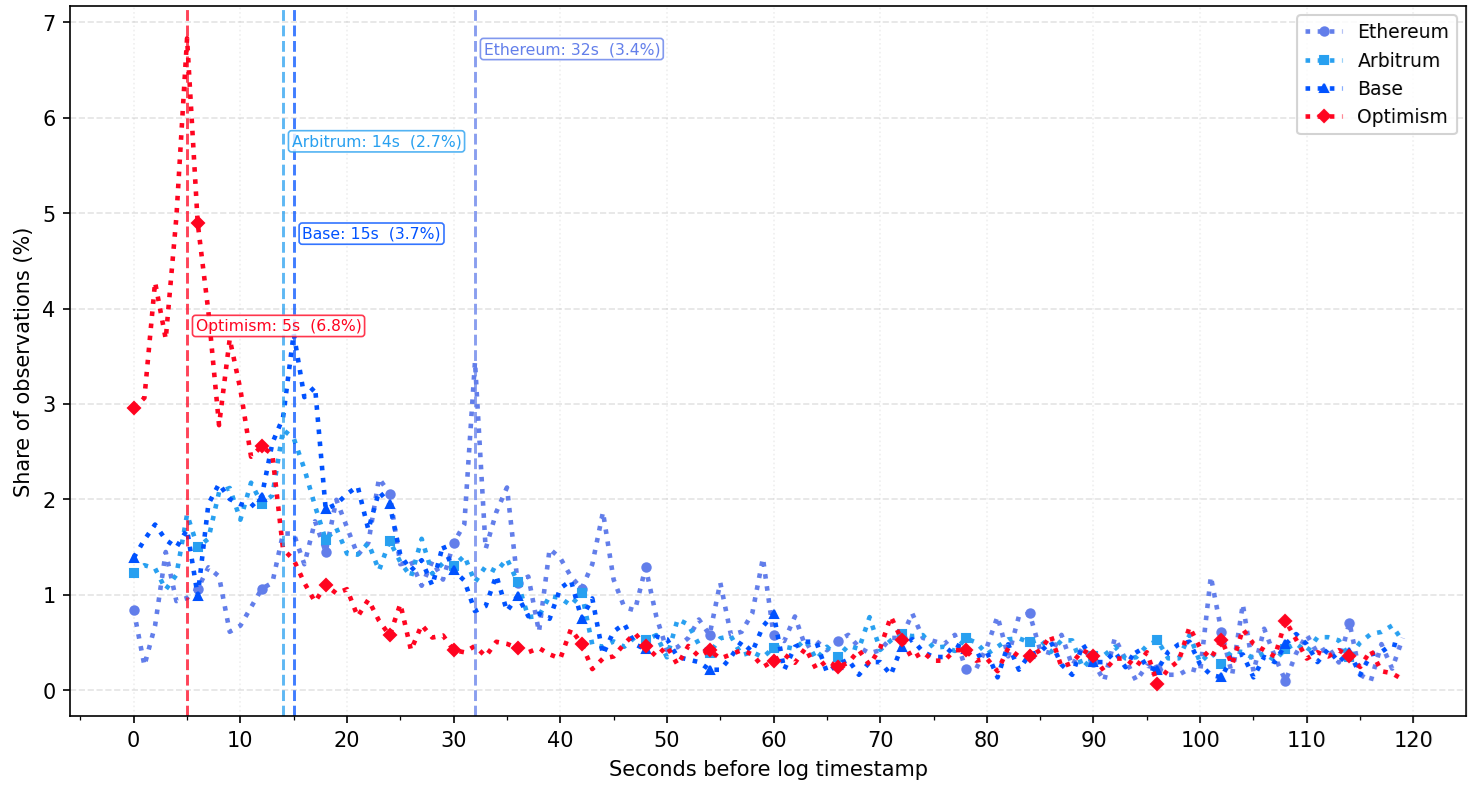}
        \caption{ETH / USD}
        \label{fig:eth_latency}
    \end{subfigure}
    \begin{subfigure}[b]{1.0\textwidth}
        \includegraphics[width=0.99\linewidth,height=5.5cm]{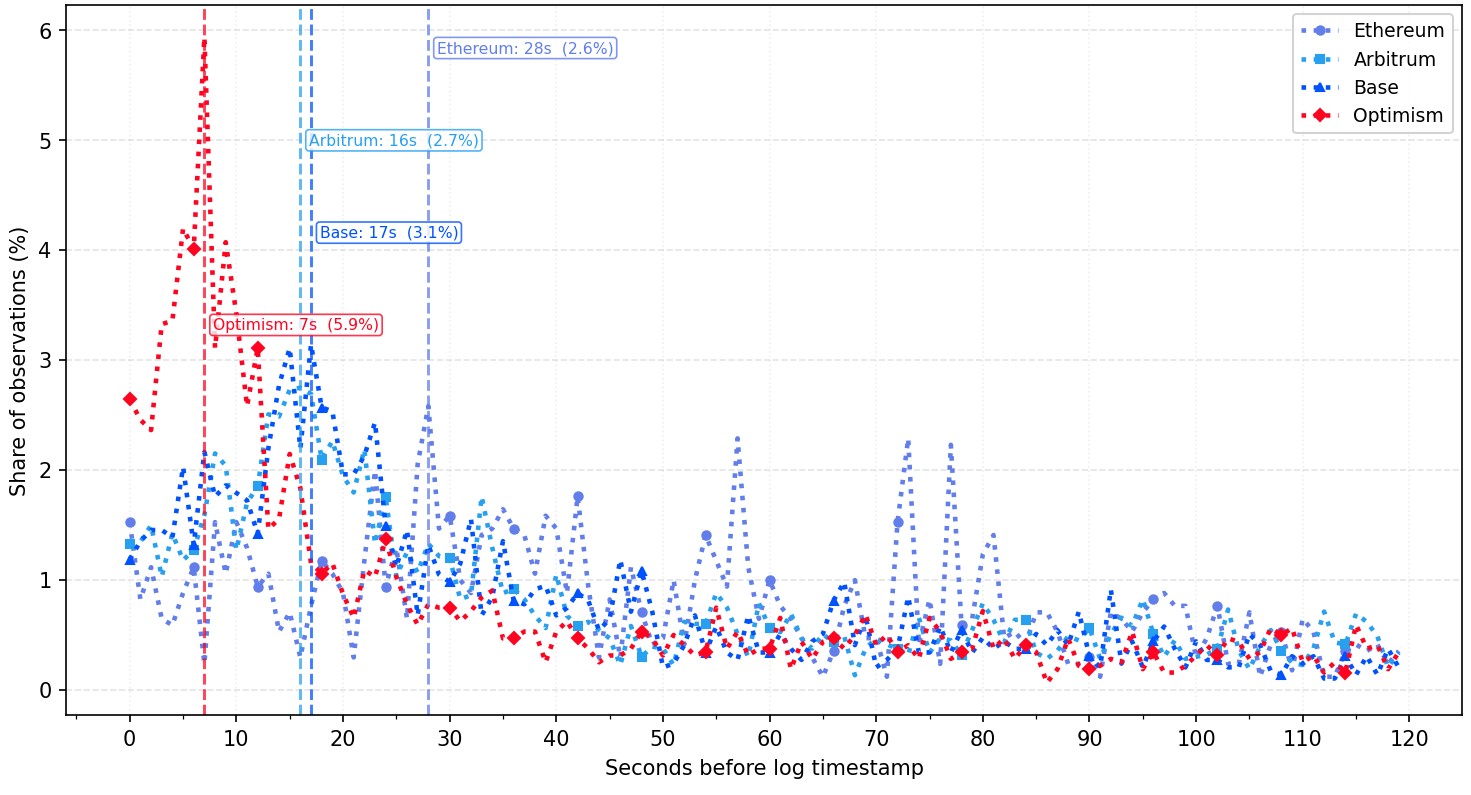}
        \caption{BTC / USD}
        \label{fig:btc_latency}
    \end{subfigure}
    \caption{Latency comparison between off-chain Binance spot prices and on-chain Chainlink oracle price observations across Arbitrum, Base, Ethereum, and Optimism.}
    \label{fig:latency_comparison}
\end{figure}

Figure~\ref{fig:latency_comparison} illustrates the distribution of matched on-chain price observations across Binance price bins for the four analyzed chains. The results suggest that Chainlink price feeds on Optimism exhibit higher temporal alignment and responsiveness to Binance spot prices compared to Arbitrum, Base, and Ethereum. In particular, price updates on Optimism appear to incorporate market movements with lower latency, indicating that price changes are reflected earlier on Optimism relative to the other chains.

\section{Exploiting Cross-Chain Latency for Speculative OEV}
\label{sec:latency}

\begin{figure}[t]
    \centering
    \includegraphics[width=0.85\linewidth]{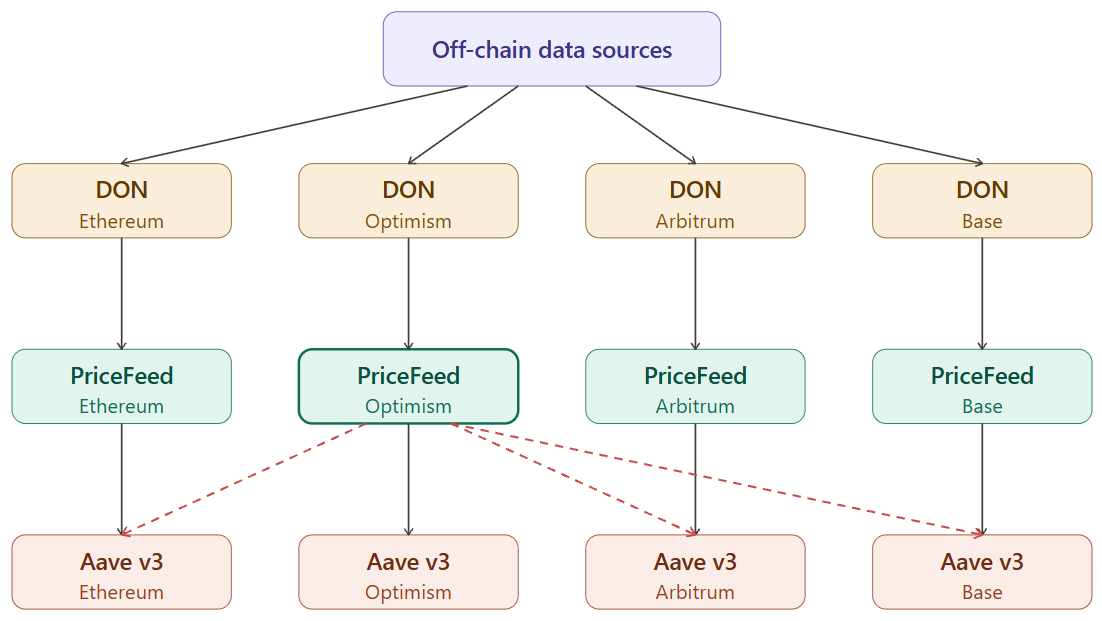}
    \caption{
    Chainlink \texttt{NewTransmission} events on Optimism act as a leading indicator for subsequent price updates on Ethereum, Arbitrum, and Base. A speculator monitoring the Optimism feed can anticipate a liquidation transaction on a target chain 
    before the corresponding oracle update is reflected on that chain. Red dashed arrows highlight the inferred speculative routing path.
    }
    \label{fig:crosschain}
\end{figure}

\begin{figure}[]
    \centering
    \includegraphics[width=0.99\linewidth,height=5cm]{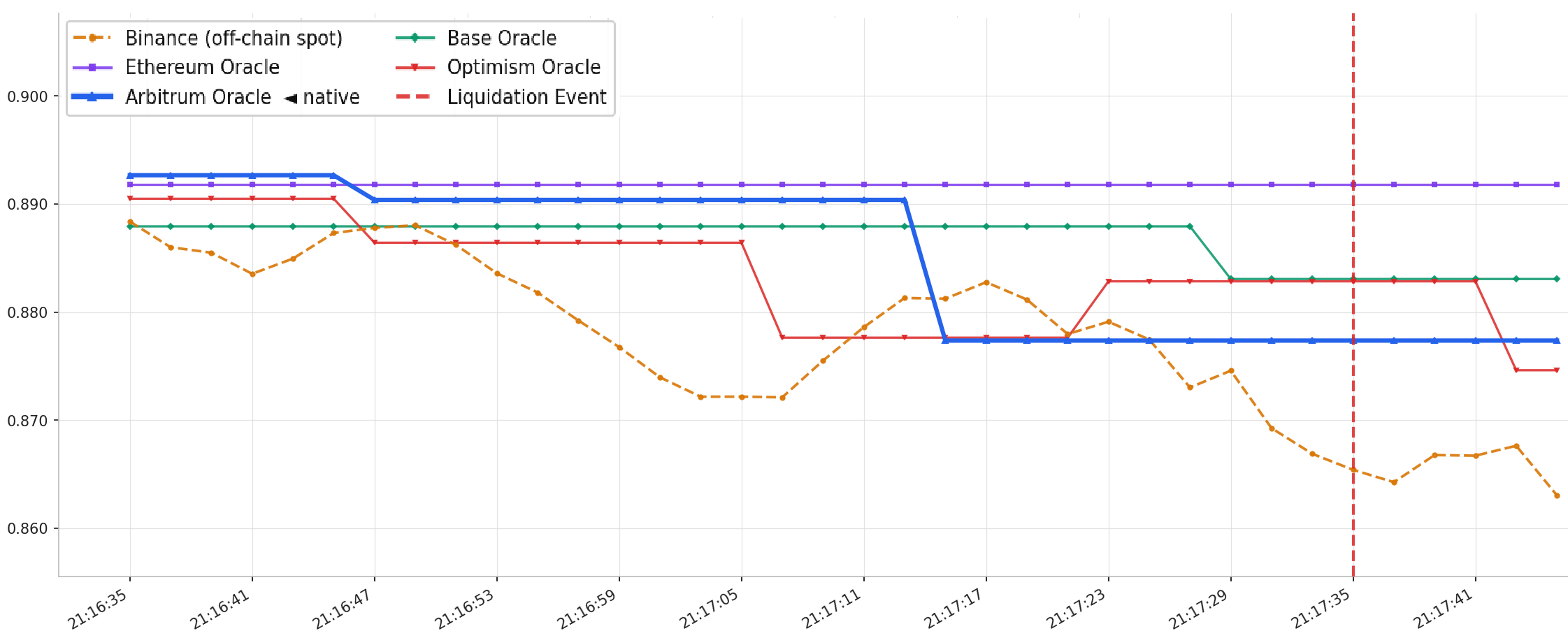}
    \caption{Health factor (HF) evolution for a BTC-collateralised Aave~V3 position
    on Arbitrum. The HF computed using Optimism price feed crosses
    the liquidation threshold ($H = 1$) at least 6 seconds before the same
    threshold is reached by the native Arbitrum Chainlink feed, defining an
    exploitable window of at least 6 seconds.}
    \label{fig:health_factor_comparison_Arbitrum_BTC}
\end{figure}

When a Chainlink feed updates on Optimism (see \figureautorefname{} \ref{fig:crosschain}), the corresponding feeds on other target chains, such as Arbitrum, may not yet have reflected the updated price. During this interval, a searcher observing Optimism can estimate the impending health factor on the target chain and, if it is expected to fall below the liquidation threshold, preemptively submit a liquidation transaction. \figureautorefname{} \ref{fig:health_factor_comparison_Arbitrum_BTC} illustrates an instance in which the Optimism price feed signaled a liquidation opportunity on Arbitrum roughly 6 seconds in advance.

\subsection{Simulation Framework}
\label{subsec:simulation}

To quantify the exploitability of the cross-chain latency gap, we replay every
historical liquidation event and determine, for each price source, the exact second at which the HF first crossed the liquidation
threshold within the observation window $[t_{\text{liq}} - 60\text{s},\;
t_{\text{liq}} + 10\text{s}]$.

\subparagraph*{Health Factor Computation.}
For each price source~$s$, the HF at time~$t$ is:
\begin{equation}
    H^{(s)}(t) \;=\;
    \frac{\text{collateral\_amount} \times P_{\text{coll}}^{(s)}(t) \times \text{LT}}
         {\text{debt\_amount} \times P_{\text{debt}}^{(s)}(t)},
    \label{eq:hf_source}
\end{equation}
where $P^{(s)}(t)$ is the last-tick-forward price from source~$s$ at time~$t$,
and $\text{LT}$ is the liquidation threshold fetched on-chain for the collateral
asset.
Stablecoin prices are fixed at 1\,USD.

\subparagraph*{Crossing-Time Detection.}
We determine the crossing times from the \texttt{NewTransmission} event timestamps.
For each event within the observation window, the running price of the affected asset is updated and the corresponding health factor $H^{(s)}$ is recomputed. The first timestamp at which $H^{(s)} \leq 1.0$ is then recorded as the crossing time $t^{(s)}_*$.

\subparagraph*{Binance Signal Filters.}
Because Binance updates every second while on-chain oracles update every
few minutes, a naive Binance signal fires frequently on noise. Two filters
are applied before a Binance crossing is accepted:
\begin{itemize}[noitemsep, topsep=4pt]
    \item \textbf{Deviation Filter:} The Binance price must differ from the
          most recent native oracle price by at least a feed- and
          chain-specific threshold $\delta_{f,c}$ (ranging from 0.05\% to
          2\%). Signals where
          $|P^{\text{Binance}} - P^{\text{native}}| / P^{\text{native}}
          < \delta_{f,c}$ are discarded as noise.
    \item \textbf{Sustain Filter:} The HF must remain at or below
          1.0 for at least 3 consecutive seconds
          before the crossing is
          accepted. This suppresses transient dips that would not trigger an
          on-chain liquidation.
\end{itemize}

\subparagraph*{Outcome Classification.}
Each liquidation window is classified into one of four outcomes, based on
whether both the reference source (Optimism or Binance) and the native
oracle crossed the threshold within the observation window:
\begin{itemize}[noitemsep, topsep=4pt]
    \item \textbf{True Positives (TP, both crossed):} Reference and native oracle both crossed
          $H \leq 1.0$. The reference signal is valid and actionable.
    \item \textbf{False Positives (FP, reference only):} The reference source crossed, but the
          native oracle never did. A bot acting on this signal would likely be unsuccessful.
\end{itemize}
From TP and FP counts we can compute the precision as $\frac{\text{TP}}{\text{TP} + \text{FP}}$.
The \emph{lead time} $\Delta t = t^{(\text{native})}_* - t^{(\text{ref})}_*$
measures the advance notice available to the searcher.
Positive $\Delta t$ means the reference source crossed first.

\subsection{Simulation Results}
\label{sec:results}

\subparagraph*{Optimism Dominance Over Native Oracle.}
Using oracle event timestamps, the Optimism oracle crossed the
liquidation threshold ($H \leq 1.0$) before the native chain oracle in the
majority of TP windows across all three target chains.
Table~\ref{tab:op_lead} reports the breakdown per chain.
In a substantially smaller fraction of TP windows, the Optimism oracle crossed
the threshold after the native oracle.
Overall, Optimism led more than twice as often as it lagged, and the average
lead was nearly double the average lag, confirming a systematic directional
advantage.

\begin{table}[t]
    \centering
    \begin{adjustbox}{max width=\columnwidth}
    \begin{tabular}{@{}lrrrrrr@{}}
        \toprule
        \textbf{Chain} & \textbf{TP} & \textbf{FP} &
        \textbf{Before (TP)} & \textbf{After (TP)} &
        \textbf{Avg Before (s)} & \textbf{Avg After (s)} \\
        \midrule
        Arbitrum & 498 &  1 & 331\ (66.5\%) & 134\ (26.9\%) & 19.40 & 10.79 \\
        Base     & 414 & 0 & 263\ (63.5\%) & 129\ (31.2\%) & 26.65 & 12.81 \\
        Ethereum & 188 &  155 & 144\ (76.6\%) & 43\ (22.9\%) & 26.95 & 8.79 \\
        \bottomrule
    \end{tabular}
    \end{adjustbox}
    \caption{Overview for HFs computed using Optimism
    price feeds versus native price feeds, for October 10, 2025. TP denotes
    windows where both sources crossed; FP denotes windows where only the
    Optimism oracle crossed. ``Before'' and ``After'' are fractions of TP
    windows. 
    }
    \label{tab:op_lead}    
\end{table}

\begin{table}
\centering
\begin{adjustbox}{max width=\columnwidth}
\begin{tabular}{@{}llrrrrrr@{}}
\toprule
\textbf{Chain} & \textbf{Price Feed} &
\textbf{Updates} & \textbf{Mean\,(s)} & \textbf{Median\,(s)} &
\textbf{Min\,(s)} & \textbf{Max\,(s)} & \textbf{Upd/hr} \\
\midrule
\multirow{7}{*}{Ethereum} & ETH / USD &  117 &  760 &  204 &  12 &  3{,}636 &  4.74 \\
         & BTC / USD  &   81 & 1,066 &  396 &  12 &  3{,}648 &  3.38 \\
         & BTC / ETH  &    8 & 7,334 & 4,680 & 228 & 23{,}364 &  0.49 \\
         & LINK / USD &  114 &  796 &  156 &  12 &  3{,}648 &  4.52 \\
         & LINK / ETH &   60 & 1,269 &   84 &  12 & 21{,}624 &  2.84 \\
         & AAVE / ETH &   25 & 3,102 &  240 &  12 & 62{,}136 &  1.16 \\
         & AAVE / USD &   92 &  956 &  168 &  12 &  3{,}636 &  3.77 \\ \midrule
\multirow{5}{*}{Arbitrum} & ETH / USD & 1{,}193 &   73 &  33 &  1 &    600 & 49.65 \\
         & ARB / USD  & 1{,}475 &   59 &  31 &  1 &    390 & 60.79 \\
         & BTC / USD  &   784 &  110 &  60 &  1 &  1{,}411 & 32.60 \\
         & LINK / USD &   513 &  169 &  60 &  1 &  2{,}160 & 21.28 \\
         & AAVE / USD &   201 &  433 &  90 &  1 &  3{,}606 &  8.32 \\
\midrule
\multirow{3}{*}{Base}     & ETH / USD &  480 &  180 &  62 &  28 &  1{,}232 & 19.96 \\
         & BTC / USD  &  444 &  194 &  90 &  18 &  1{,}232 & 18.52 \\
         & AAVE / USD &  185 &  422 &  60 &  26 &  9{,}486 &  8.53 \\
\midrule
\multirow{5}{*}{Optimism} & OP / USD   &  689 &  135 &  44 &   2 &  1{,}864 & 26.61 \\
         & ETH / USD &  611 &  148 &  44 &   2 &  1{,}934 & 24.39 \\
         & LINK / USD &  609 &  147 &  44 &   2 &  1{,}866 & 24.54 \\
         & AAVE / USD &  595 &  158 &  44 &   2 &  1{,}846 & 22.72 \\
         & BTC / USD  &  530 &  171 &  54 &   2 &  1{,}928 & 21.05 \\
\bottomrule
\end{tabular}
\end{adjustbox}
\caption{Oracle update frequency by feed and chain (inter-event intervals) on October 10, 2025.}
\label{tab:update_freq_chain}
\end{table}

\subparagraph*{Oracle Update Frequency.}
\label{subsec:update_freq}
Table~\ref{tab:update_freq_chain} reports inter-event update intervals
computed from \texttt{AnswerUpdated} timestamps across all observed feeds.
Optimism oracles update every 181\,s on average, substantially
faster than Ethereum for instance ($\approx$1{,}466\,s), which directly explains the
directional lead observed above. Ethereum oracles update roughly 2-3~times
per hour, whereas Base and Arbitrum oracles update approximately
13-17~times per hour.

\subparagraph*{Cross-Chain Speculative OEV.} Our results show that oracle updates observed on Optimism can serve as predictive signals for forthcoming price updates on other chains. However, because the mempools of L2 networks such as Arbitrum and Base are largely opaque, successful liquidation execution remains probabilistic. A baseline speculative strategy therefore consists of repeatedly broadcasting liquidation transactions for positions expected to become undercollateralized following a corresponding oracle update on the target chain.

We compute the success rate as the fraction of cases where the native oracle update occurs within a maximum window of $T$ seconds after the corresponding Optimism update, implying that a speculative liquidator would continuously emit liquidation transactions during this interval. Operational costs are estimated assuming one transaction per block, with transaction emission stopping once a liquidation transaction is included in the same block as the native oracle update. Consequently, costs correspond to the number of blocks between the observed Optimism update and the subsequent target-chain update.
We evaluate these metrics across all observed price updates and report average success rates and costs in Table~\ref{tab:baseline_per_feed}. The highest success rate on Arbitrum is observed for liquidations linked to the ETH/USD price feed, reaching an average of 96.7\% while requiring the emission of approximately one transaction across 99.8 consecutive blocks on average.

\begin{table}[]
\centering
\begin{adjustbox}{max width=\columnwidth}
\begin{tabular}{l l r r r r r r r r r r r}
\toprule
& & & \multicolumn{2}{c}{\textbf{$T = 15$\,s}}
  & \multicolumn{2}{c}{\textbf{$T = 30$\,s}}
  & \multicolumn{2}{c}{\textbf{$T = 60$\,s}}
  & \multicolumn{2}{c}{\textbf{$T = 90$\,s}}
  & \multicolumn{2}{c}{\textbf{$T = 120$\,s}} \\
\cmidrule(lr){4-5}\cmidrule(lr){6-7}\cmidrule(lr){8-9}
\cmidrule(lr){10-11}\cmidrule(lr){12-13}
\textbf{Chain} & \textbf{Price Feed} & \textbf{$n$}
  & \textbf{SR} & \textbf{Cost}
  & \textbf{SR} & \textbf{Cost}
  & \textbf{SR} & \textbf{Cost}
  & \textbf{SR} & \textbf{Cost}
  & \textbf{SR} & \textbf{Cost} \\
\midrule
\multirow{4}{*}{Arbitrum}
  & ETH  / USD & 591 & 41.9 &  48.7 &  81.5 &  71.4 &  92.3 &  87.3 &  95.1 &  94.9 &  96.7 &  99.8  \\
  & BTC  / USD & 484 & 36.8 &  49.5 &  67.0 &  79.0 &  82.6 & 110.5 &  89.8 & 127.5 &  91.3 & 139.2  \\
  & LINK / USD & 495 & 27.5 &  52.2 &  53.1 &  87.9 &  68.2 & 134.2 &  77.9 & 167.6 &  82.4 & 191.5  \\
  & AAVE / USD & 304 & 12.7 &  56.5 &  25.3 & 105.3 &  38.4 & 188.2 &  43.7 & 259.3 &  52.1 & 322.0 \\
\midrule
\multirow{3}{*}{Base}
  & ETH  / USD & 500 & 20.9 &   7.2 &  47.3 &  11.8 &  68.1 &  18.0 &  78.7 &  22.2 &  81.8 &  25.1  \\
  & BTC  / USD & 413 & 28.1 &   7.0 &  53.5 &  11.0 &  69.1 &  16.8 &  76.6 &  21.0 &  80.7 &  24.3  \\
  & AAVE / USD & 328 & 15.4 &   7.4 &  33.0 &  12.7 &  46.1 &  21.8 &  53.1 &  29.3 &  56.2 &  36.1  \\
\bottomrule
\end{tabular}
\end{adjustbox}
\caption{Success rate (SR, \%) and operational cost (avg.\ transactions) of the baseline
speculative liquidator strategy under continuous broadcast over window $[0, T]$.
$n$ denotes the number of
matched Optimism price update events per feed. 
$[0,T]$ is charged. 
}
\label{tab:baseline_per_feed}
\end{table}

\subparagraph*{Transaction Distance Optimization.}

Ideally, a liquidation should be placed immediately after the oracle update to maximize the probability of capturing the opportunity before competing searchers.
Our analysis of priority fees used by Chainlink transmitters suggests that oracle update transactions target timely inclusion rather than top-of-block placement, with fees closely tracking the historical median (50th percentile) of preceding blocks.

To evaluate this behavior, we used the \texttt{eth\_feeHistory} RPC method to retrieve 50th percentile priority fees for up to 100 blocks preceding each oracle update transaction. Based on rolling windows of 1–100 prior blocks, we computed median fee estimates and compared them against the actual priority fees paid by oracle update transactions. We define the success rate as the fraction of cases where the estimated fee was lower than the oracle transaction fee while still exceeding the fee of the last transaction included in the block, indicating that a liquidation transaction could have been included in the same block after the oracle update. We additionally compute the exact match ratio, capturing cases where the estimated and observed priority fees were identical.

Since Arbitrum follows a first-come, first-served ordering policy, priority fees have limited influence and transaction ordering is primarily latency-driven. In contrast, Base prioritizes transactions with higher priority gas fees. We find that our approach achieves high same-block inclusion success rates on Base across the three analyzed price feeds (see \tableautorefname{} \ref{tab:transaction_distance} in \appendixautorefname{} \ref{sec:appendix_transaction_distance_optimization}). For ETH / USD, for example, using the median 50th percentile priority fee over the previous 37 blocks yields a 94.48\% success rate and a 7.64\% exact match ratio. The results further indicate a trade-off between maximizing exact fee matches and achieving higher inclusion success rates.

\section{Conclusion}
\label{sec:conclusion}

We presented the first empirical analysis of speculative Oracle Extractable Value (OEV) in Aave liquidations across Layer-2 blockchains. Using a novel detection methodology, we identified 64 speculative liquidators on Arbitrum, Base, and Optimism, representing 57\% of all detected liquidators, as well as 831 successful speculative liquidations, accounting for 39\% of all successful detected liquidations during our observation period on October 10, 2025.
Our analysis further demonstrated that cross-chain latency asymmetries in Chainlink oracle updates create statistically predictable exploitation windows. Although independent DONs consume largely identical off-chain price information nearly simultaneously, differences in oracle configurations and block times cause updates to be published at different times across chains. In particular, we showed that oracle updates on Optimism can act as predictive signals for subsequent updates on Arbitrum and Base, enabling speculative cross-chain OEV extraction.
These findings suggest that the structural conditions enabling speculative cross-chain OEV are persistent and may become increasingly relevant as Layer-2 adoption grows. At the same time, our study currently focuses on a single intraday observation window, implying that the measured activity likely represents solely a lower bound on the true prevalence of speculative OEV.

\newpage
\bibliography{references}

@article{Qin2021,
  author    = {Qin, Kaihua 
  and Zhou, Liyi 
  and Gamito, Pablo 
  and Jovanovic, Philipp 
  and Gervais, Arthur},
  title     = {An Empirical Study of {DeFi} Liquidations: Incentives, Risks, and Instabilities},
  journal   = {Proceedings of the ACM Internet Measurement Conference (IMC)},
  year      = {2021},
  doi       = {10.1145/3487552.3487811},
}

@misc{Greene2024,
  author    = {Greene, Jacob and {API3 DAO}},
  title     = {Oracle Extractable Value ({OEV}) through Order Flow Auctions},
  year      = {2024},
  note      = {API3 Whitepaper. Available at \url{https://github.com/api3dao/oev-litepaper}},
}

@inproceedings{Torres2025,
  author    = {Torres, Christof Ferreira and others},
  title     = {Rolling in the Shadows: Analyzing the Extraction of {MEV} Across Layer-2 Rollups},
  booktitle = {arXiv preprint arXiv:2405.00138},
  year      = {2025},
}

@misc{Finkbeiner2025,
  author    = {Finkbeiner, Colin and Almashaqbeh, Ghada},
  title     = {{SoK}: Blockchain Oracles Between Theory and Practice},
  year      = {2025},
  note      = {ePrint 2025/2106. Available at \url{https://eprint.iacr.org/2025/2106}},
}

@misc{Humphry2024,
  author    = {Humphry, Rebecca and others},
  title     = {Review of Maximal Extractable Value \& Blockchain Oracles},
  year      = {2024},
  note      = {FCA Research Note},
}

@inproceedings{Yang2024,
  author    = {Yang, Sen
  and Zhang, Fan
  and Ken, Huang
  and Chen, Xi
  and Yang, Youwei
  and Zhu, Feng},
  title     = {SoK: MEV Countermeasures: Theory and Practice},
  booktitle = {DeFi '24: Proceedings of the Workshop on Decentralized Finance and Security},
  year      = {2024},
  doi       = {10.1145/3689931.3694911},
}

@inproceedings{PriceOracle2025,
  author    = {Gansäuer, Robin
  and Ben Aoun, Hichem
  and Droll, Jan
  and Hartenstein, Hannes},
  title     = {Price Oracle Accuracy Across Blockchains: A Measurement and Analysis},
  booktitle = {Financial Cryptography and Data Security. FC 2025 International Workshops},
  year      = {2025},
  doi       = {10.1007/978-3-032-00492-5_3},
}

@article{Nadler2026,
author = {Nadler, Matthias
and Schuler, Katrin 
and Schär, Fabian},
title = {Blockchain price oracles: Accuracy and violation recovery},
journal = {Journal of Corporate Finance},
volume = {96},
pages = {102908},
year = {2026},
issn = {0929-1199},
doi = {https://doi.org/10.1016/j.jcorpfin.2025.102908},
}

@INPROCEEDINGS{Vakhmyanin2023,
  author={Vakhmyanin, Ivan
  and Volkovich, Yana},
  booktitle={2023 IEEE International Conference on Blockchain and Cryptocurrency (ICBC)}, 
  title={Price Arbitrage for DeFi Derivatives}, 
  year={2023},
  volume={},
  number={},
  pages={1-4},
  doi={10.1109/ICBC56567.2023.10174884}
  }

@InProceedings{Andreoulis2026,
author="Andreoulis, Nikolaos
and Maggio, Marco Di
and Merino, Louis-Henri
and Montag, Kyle
and Ward, James",
editor="Leonardos, Stefanos
and Goharshady, Amir K.
and Knottenbelt, William
and Pardalos, Panos",
title="Designing for Fair Oracle Extractable Value: A Theoretical Framework and Empirical Findings from DeFi",
booktitle="Mathematical Research for Blockchain Economy",
year="2026",
publisher="Springer Nature Switzerland",
address="Cham",
pages="48--62",
isbn="978-3-032-13377-9"
}

@misc{Gogol2024,
title={Cross-Rollup MEV: Non-Atomic Arbitrage Across L2 Blockchains}, 
author={Gogol, Krzysztof
and Messias, Johnnatan
and Miori, Deborah
and Tessone, Claudio
and Livshits, Benjamin},
year={2024},
eprint={2406.02172},
archivePrefix={arXiv},
primaryClass={cs.CR},
url={https://arxiv.org/abs/2406.02172}, 
}

@inproceedings{Solmaz2025OptimisticMEV,
  author    = {Ozan Solmaz 
  and Lioba Heimbach 
  and Yann Vonlanthen 
  and Roger Wattenhofer},
  title     = {Optimistic MEV in Ethereum Layer 2s: Why Blockspace Is Always in Demand},
  booktitle = {Proceedings of the 7th Conference on Advances in Financial Technologies (AFT 2025)},
  series    = {Leibniz International Proceedings in Informatics (LIPIcs)},
  year      = {2025},
  publisher = {Schloss Dagstuhl -- Leibniz-Zentrum f{\"u}r Informatik},
  doi       = {10.4230/LIPIcs.AFT.2025.28},
  url       = {https://doi.org/10.4230/LIPIcs.AFT.2025.28}
}

@misc{coingecko_api,
  author       = {{CoinGecko}},
  title        = {CoinGecko API Documentation},
  year         = {2025},
  howpublished = {\url{https://www.coingecko.com/en/api}},
  note         = {Accessed: 2025-05-24}
}

\appendix

\section{Liquidator Profits and Costs}
\label{sec:appendix_liquidator_profits_costs}

\begin{table}[H]
    \centering
    \begin{adjustbox}{max width=\columnwidth}
    \begin{tabular}{l c r r r r r}
        \toprule
        & & \multicolumn{5}{c}{\textbf{Profit (USD)}} \\
        \cline{3-7}
        \textbf{Chain} & \textbf{Liquidator} & \textbf{Total} & \textbf{Min} & \textbf{Mean} & \textbf{Median} & \textbf{Max} \\
        \midrule
        \multirow{10}{*}{Arbitrum}
        & \cellcolor{gray!20}\href{https://arbiscan.io/address/0xb9516c655831917704A5000bA6f4EB64E816DC2d}{\texttt{0xb9516c655831917704A5000bA6f4EB64E816DC2d}} & \cellcolor{gray!20}313,737.26 & \cellcolor{gray!20}-2,536.95 & \cellcolor{gray!20}2,752.08 & \cellcolor{gray!20}178.69 & \cellcolor{gray!20}78,688.71 \\
        & \href{https://arbiscan.io/address/0x001f8151FC6d0a14608B48F1d2a2AeA66c80cF38}{\texttt{0x001f8151FC6d0a14608B48F1d2a2AeA66c80cF38}} & 223,092.53 & -163.48 & 1,582.22 & 387.74 & 40,502.66 \\
        & \href{https://arbiscan.io/address/0x48daab9f7Ed6E3184c26D9dDACcc356D52D6237F}{\texttt{0x48daab9f7Ed6E3184c26D9dDACcc356D52D6237F}} & 335,674.81 & -524.72 & 3,571.01 & 1,749.04 & 37,264.96 \\
        & \cellcolor{gray!20}\href{https://arbiscan.io/address/0xd249aB6aaC55a7AD0ceF7fCFD672d47387D7e70F}{\texttt{0xd249aB6aaC55a7AD0ceF7fCFD672d47387D7e70F}} & \cellcolor{gray!20}101,111.82 & \cellcolor{gray!20}-2,822.44 & \cellcolor{gray!20}1,579.87 & \cellcolor{gray!20}464.55 & \cellcolor{gray!20}27,635.05 \\
        & \cellcolor{gray!20}\href{https://arbiscan.io/address/0x6290c280f1393d33f04d6A59993cbe7d3ECCDFcF}{\texttt{0x6290c280f1393d33f04d6A59993cbe7d3ECCDFcF}} & \cellcolor{gray!20}94,106.90 & \cellcolor{gray!20}-2,150.06 & \cellcolor{gray!20}1,845.23 & \cellcolor{gray!20}301.15 & \cellcolor{gray!20}20,860.18 \\
        & \href{https://arbiscan.io/address/0x545875b6975eb527726C87B82B5f031ED861De8A}{\texttt{0x545875b6975eb527726C87B82B5f031ED861De8A}} & 915,453.18 & -20.77 & 15,516.16 & 818.18 & 488,634.40 \\
        & \cellcolor{gray!20}\href{https://arbiscan.io/address/0x9f83664976aDc0abe9C375151B0E5B36eE7703D7}{\texttt{0x9f83664976aDc0abe9C375151B0E5B36eE7703D7}} & \cellcolor{gray!20}1,379.63 & \cellcolor{gray!20}-13.25 & \cellcolor{gray!20}26.53 & \cellcolor{gray!20}19.17 & \cellcolor{gray!20}150.50 \\
        & \href{https://arbiscan.io/address/0xa441ed20E5A5c51C067549316B86F84CC988D284}{\texttt{0xa441ed20E5A5c51C067549316B86F84CC988D284}} & 259,371.48 & 388.17 & 9,606.35 & 4,364.83 & 83,620.76 \\
        & \href{https://arbiscan.io/address/0x4044E9E4E9d0ef73026106d097DCdB1d609436A7}{\texttt{0x4044E9E4E9d0ef73026106d097DCdB1d609436A7}} & 83,023.14 & -1,136.24 & 5,188.95 & 2,416.98 & 27,016.66 \\
        & \href{https://arbiscan.io/address/0x4Ae48f73Abc9F6f7A918ff405584B9314A5df0b8}{\texttt{0x4Ae48f73Abc9F6f7A918ff405584B9314A5df0b8}} & 186,776.46 & 1,012.07 & 9,830.34 & 9,887.57 & 24,661.42 \\
        \midrule
        \multirow{10}{*}{Base}
        & \href{https://basescan.org/address/0xD251c1325c5d7b29C6219912D8648a3149cDF57B}{\texttt{0xD251c1325c5d7b29C6219912D8648a3149cDF57B}} & 85,695.53 & 11.28 & 856.96 & 261.98 & 14,264.82 \\
        & \cellcolor{gray!20}\href{https://basescan.org/address/0xc89c328609aB58E256Cd2b5aB4F4aF2EFb9fcA33}{\texttt{0xc89c328609aB58E256Cd2b5aB4F4aF2EFb9fcA33}} & \cellcolor{gray!20}184,351.11 & \cellcolor{gray!20}0.18 & \cellcolor{gray!20}3,478.32 & \cellcolor{gray!20}395.27 & \cellcolor{gray!20}102,233.99 \\
        & \cellcolor{gray!20}\href{https://basescan.org/address/0x3ba19cB44eF72B0B198325E17623BcE10Bae2753}{\texttt{0x3ba19cB44eF72B0B198325E17623BcE10Bae2753}} & \cellcolor{gray!20}2,746.95 & \cellcolor{gray!20}-49.52 & \cellcolor{gray!20}124.86 & \cellcolor{gray!20}23.19 & \cellcolor{gray!20}941.10 \\
        & \href{https://basescan.org/address/0x964AeE3e4E3BBc7245B33dA097030e95EE408170}{\texttt{0x964AeE3e4E3BBc7245B33dA097030e95EE408170}} & 13,752.00 & 9.89 & 416.73 & 184.58 & 3,275.35 \\
        & \cellcolor{gray!20}\href{https://basescan.org/address/0xec5aCd2dfdf3B2e4dcb955C7eC8B2b74605d8E9c}{\texttt{0xec5aCd2dfdf3B2e4dcb955C7eC8B2b74605d8E9c}} & \cellcolor{gray!20}86,078.39 & \cellcolor{gray!20}-674.62 & \cellcolor{gray!20}7,173.20 & \cellcolor{gray!20}367.41 & \cellcolor{gray!20}43,622.09 \\
        & \cellcolor{gray!20}\href{https://basescan.org/address/0x888888887A487f209e31a692B227d8D1ff9070ba}{\texttt{0x888888887A487f209e31a692B227d8D1ff9070ba}} & \cellcolor{gray!20}1,271.19 & \cellcolor{gray!20}-1.49 & \cellcolor{gray!20}57.78 & \cellcolor{gray!20}15.86 & \cellcolor{gray!20}481.75 \\
        & \href{https://basescan.org/address/0x3347277366deCC91d65D2762E792E6BaA471b805}{\texttt{0x3347277366deCC91d65D2762E792E6BaA471b805}} & 92,120.15 & 233.61 & 4,386.67 & 2,528.14 & 43,800.50 \\
        & \href{https://basescan.org/address/0xaDDD8ECec572E3BA9975578e1927221eC25Dab50}{\texttt{0xaDDD8ECec572E3BA9975578e1927221eC25Dab50}} & 12,510.15 & 0.13 & 695.01 & 100.11 & 4,437.22 \\
        & \href{https://basescan.org/address/0x3f482cA03E6d5FA63884918d5dC379F14cA0aD86}{\texttt{0x3f482cA03E6d5FA63884918d5dC379F14cA0aD86}} & 40,352.58 & 117.56 & 2,373.68 & 1,408.65 & 6,433.55 \\
        & \href{https://basescan.org/address/0xf4Cd829526070C2C0af9E6Dce2C9255130f7ec68}{\texttt{0xf4Cd829526070C2C0af9E6Dce2C9255130f7ec68}} & 9,816.33 & 0.04 & 701.17 & 24.37 & 5,949.76 \\
        \midrule
        \multirow{10}{*}{Optimism}
        & \href{https://optimistic.etherscan.io/address/0x7518Ba8b4021D83caFB1e46Fa1250f54fA86b6Ab}{\texttt{0x7518Ba8b4021D83caFB1e46Fa1250f54fA86b6Ab}} & 222,090.59 & -120.87 & 1,314.15 & 394.28 & 16,743.39 \\
        & \cellcolor{gray!20}\href{https://optimistic.etherscan.io/address/0x8888888885FcA4862619cAFeda9b03768C930012}{\texttt{0x8888888885FcA4862619cAFeda9b03768C930012}} & \cellcolor{gray!20}70,688.68 & \cellcolor{gray!20}0.08 & \cellcolor{gray!20}631.15 & \cellcolor{gray!20}65.51 & \cellcolor{gray!20}8,503.09 \\
        & \href{https://optimistic.etherscan.io/address/0x4661f836F07c78a01D5490bFf6e8587B9291C6B8}{\texttt{0x4661f836F07c78a01D5490bFf6e8587B9291C6B8}} & 182,296.26 & -71.45 & 2,048.27 & 248.68 & 20,609.11 \\
        & \href{https://optimistic.etherscan.io/address/0x4db5FBb9246F66bd83661D680F308cc05c5FC3Bf}{\texttt{0x4db5FBb9246F66bd83661D680F308cc05c5FC3Bf}} & 77,547.49 & 0.02 & 881.22 & 519.23 & 9,038.31 \\
        & \href{https://optimistic.etherscan.io/address/0x874Bd329Adf526A61FeE5766B9E96c9519aeBb1b}{\texttt{0x874Bd329Adf526A61FeE5766B9E96c9519aeBb1b}} & 33,819.16 & -1.42 & 439.21 & 21.54 & 19,703.41 \\
        & \href{https://optimistic.etherscan.io/address/0x1dC68046E71429F646Dd6B2d0c02385aE61c2129}{\texttt{0x1dC68046E71429F646Dd6B2d0c02385aE61c2129}} & 23,216.09 & 48.50 & 438.04 & 330.72 & 3,625.10 \\
        & \cellcolor{gray!20}\href{https://optimistic.etherscan.io/address/0x083CfA7FD187Be983ce5D519fE7ae78357779998}{\texttt{0x083CfA7FD187Be983ce5D519fE7ae78357779998}} & \cellcolor{gray!20}50,537.69 & \cellcolor{gray!20}-107.67 & \cellcolor{gray!20}1,203.28 & \cellcolor{gray!20}949.51 & \cellcolor{gray!20}5,728.92 \\
        & \href{https://optimistic.etherscan.io/address/0x59c5f1B4FF4fdf04729787EeE84DdBDf787A3033}{\texttt{0x59c5f1B4FF4fdf04729787EeE84DdBDf787A3033}} & 420.99 & -0.00 & 11.38 & 0.03 & 195.32 \\
        & \href{https://optimistic.etherscan.io/address/0x4B1142124c661B48aD7F98d8C7C8db03016B1F88}{\texttt{0x4B1142124c661B48aD7F98d8C7C8db03016B1F88}} & 16,857.15 & 148.55 & 648.35 & 493.21 & 2,687.90 \\
        & \href{https://optimistic.etherscan.io/address/0xe3BadF4DCf5908472468e27CF6b438f3F11D9c5D}{\texttt{0xe3BadF4DCf5908472468e27CF6b438f3F11D9c5D}} & 92,548.48 & 71.76 & 3,559.56 & 607.31 & 31,731.91 \\
        \bottomrule
    \end{tabular}
    \end{adjustbox}
    \caption{Overview of the top 10 liquidators across Arbitrum, Base, and Optimism, ranked by the number of liquidations performed, including total realized profit as well as minimum, average, median, and maximum profit per liquidation.}
    \label{tab:liquidator_profits}
\end{table}

\tableautorefname{} \ref{tab:liquidator_profits} and \tableautorefname{} \ref{tab:liquidator_costs} provide an overview of the realized profits and incurred costs for the top 10 liquidators. All monetary values are denominated in USD, where reported profits correspond to net profits, i.e., gross profits after deducting transaction costs. Profits are calculated by computing the USD-denominated value of the collateral received minus the USD-denominated value of the debt asset repaid to Aave. Token prices in USD were obtained via the CoinGecko API \cite{coingecko_api} using market data for October 10, 2025.
Transaction costs are computed by aggregating the fees of all identical transactions identified within the 10 preceding and 10 succeeding blocks surrounding a liquidation transaction, including the fee of the liquidation transaction itself. Transaction fees are calculated as the product of gas used and the gas price specified by the transaction sender, with the resulting ETH-denominated value subsequently converted to USD using the ETH/USD exchange rate provided by CoinGecko for October 10, 2025.

\begin{table}[H]
    \centering
    \begin{adjustbox}{max width=\columnwidth}
    \begin{tabular}{l c r r r r r}
        \toprule
        & & \multicolumn{5}{c}{\textbf{Cost (USD)}} \\
        \cline{3-7}
        \textbf{Chain} & \textbf{Liquidator} & \textbf{Total} & \textbf{Min} & \textbf{Mean} & \textbf{Median} & \textbf{Max} \\
        \midrule
        \multirow{10}{*}{Arbitrum}
        & \cellcolor{gray!20}\href{https://arbiscan.io/address/0xb9516c655831917704A5000bA6f4EB64E816DC2d}{\texttt{0xb9516c655831917704A5000bA6f4EB64E816DC2d}} & \cellcolor{gray!20}53,692.67 & \cellcolor{gray!20}0.11 & \cellcolor{gray!20}470.99 & \cellcolor{gray!20}172.68 & \cellcolor{gray!20}2,692.61 \\
        & \href{https://arbiscan.io/address/0x001f8151FC6d0a14608B48F1d2a2AeA66c80cF38}{\texttt{0x001f8151FC6d0a14608B48F1d2a2AeA66c80cF38}} & 21,382.83 & 15.28 & 151.65 & 135.78 & 752.49 \\
        & \href{https://arbiscan.io/address/0x48daab9f7Ed6E3184c26D9dDACcc356D52D6237F}{\texttt{0x48daab9f7Ed6E3184c26D9dDACcc356D52D6237F}} & 6,606.56 & 7.80 & 70.28 & 61.81 & 271.32 \\
        & \cellcolor{gray!20}\href{https://arbiscan.io/address/0xd249aB6aaC55a7AD0ceF7fCFD672d47387D7e70F}{\texttt{0xd249aB6aaC55a7AD0ceF7fCFD672d47387D7e70F}} & \cellcolor{gray!20}32,872.55 & \cellcolor{gray!20}8.82 & \cellcolor{gray!20}513.63 & \cellcolor{gray!20}232.65 & \cellcolor{gray!20}2,762.73 \\
        & \cellcolor{gray!20}\href{https://arbiscan.io/address/0x6290c280f1393d33f04d6A59993cbe7d3ECCDFcF}{\texttt{0x6290c280f1393d33f04d6A59993cbe7d3ECCDFcF}} & \cellcolor{gray!20}35,062.21 & \cellcolor{gray!20}13.29 & \cellcolor{gray!20}687.49 & \cellcolor{gray!20}221.14 & \cellcolor{gray!20}2,853.53 \\
        & \href{https://arbiscan.io/address/0x545875b6975eb527726C87B82B5f031ED861De8A}{\texttt{0x545875b6975eb527726C87B82B5f031ED861De8A}} & 6,371.32 & 0.02 & 107.99 & 101.28 & 432.70 \\
        & \cellcolor{gray!20}\href{https://arbiscan.io/address/0x9f83664976aDc0abe9C375151B0E5B36eE7703D7}{\texttt{0x9f83664976aDc0abe9C375151B0E5B36eE7703D7}} & \cellcolor{gray!20}480.26 & \cellcolor{gray!20}7.30 & \cellcolor{gray!20}9.24 & \cellcolor{gray!20}8.86 & \cellcolor{gray!20}15.54 \\
        & \href{https://arbiscan.io/address/0xa441ed20E5A5c51C067549316B86F84CC988D284}{\texttt{0xa441ed20E5A5c51C067549316B86F84CC988D284}} & 3,490.21 & 27.81 & 129.27 & 125.45 & 342.14 \\
        & \href{https://arbiscan.io/address/0x4044E9E4E9d0ef73026106d097DCdB1d609436A7}{\texttt{0x4044E9E4E9d0ef73026106d097DCdB1d609436A7}} & 7,083.42 & 77.06 & 442.71 & 219.54 & 1,798.48 \\
        & \href{https://arbiscan.io/address/0x4Ae48f73Abc9F6f7A918ff405584B9314A5df0b8}{\texttt{0x4Ae48f73Abc9F6f7A918ff405584B9314A5df0b8}} & 8,645.06 & 85.17 & 455.00 & 334.47 & 1,264.70 \\
        \midrule
        \multirow{10}{*}{Base}
        & \href{https://basescan.org/address/0xD251c1325c5d7b29C6219912D8648a3149cDF57B}{\texttt{0xD251c1325c5d7b29C6219912D8648a3149cDF57B}} & 2,477.98 & 3.37 & 24.78 & 20.26 & 237.61 \\
        & \cellcolor{gray!20}\href{https://basescan.org/address/0xc89c328609aB58E256Cd2b5aB4F4aF2EFb9fcA33}{\texttt{0xc89c328609aB58E256Cd2b5aB4F4aF2EFb9fcA33}} & \cellcolor{gray!20}22,483.21 & \cellcolor{gray!20}3.40 & \cellcolor{gray!20}424.21 & \cellcolor{gray!20}32.60 & \cellcolor{gray!20}7,134.96 \\
        & \cellcolor{gray!20}\href{https://basescan.org/address/0x3ba19cB44eF72B0B198325E17623BcE10Bae2753}{\texttt{0x3ba19cB44eF72B0B198325E17623BcE10Bae2753}} & \cellcolor{gray!20}295.57 & \cellcolor{gray!20}0.26 & \cellcolor{gray!20}13.43 & \cellcolor{gray!20}5.49 & \cellcolor{gray!20}64.84 \\
        & \href{https://basescan.org/address/0x964AeE3e4E3BBc7245B33dA097030e95EE408170}{\texttt{0x964AeE3e4E3BBc7245B33dA097030e95EE408170}} & 591.82 & 1.40 & 17.93 & 12.00 & 127.40 \\
        & \cellcolor{gray!20}\href{https://basescan.org/address/0xec5aCd2dfdf3B2e4dcb955C7eC8B2b74605d8E9c}{\texttt{0xec5aCd2dfdf3B2e4dcb955C7eC8B2b74605d8E9c}} & \cellcolor{gray!20}4,590.99 & \cellcolor{gray!20}40.77 & \cellcolor{gray!20}382.58 & \cellcolor{gray!20}354.35 & \cellcolor{gray!20}931.60 \\
        & \cellcolor{gray!20}\href{https://basescan.org/address/0x888888887A487f209e31a692B227d8D1ff9070ba}{\texttt{0x888888887A487f209e31a692B227d8D1ff9070ba}} & \cellcolor{gray!20}112.61 & \cellcolor{gray!20}0.44 & \cellcolor{gray!20}5.12 & \cellcolor{gray!20}5.82 & \cellcolor{gray!20}10.58 \\
        & \href{https://basescan.org/address/0x3347277366deCC91d65D2762E792E6BaA471b805}{\texttt{0x3347277366deCC91d65D2762E792E6BaA471b805}} & 6,737.97 & 25.51 & 320.86 & 63.46 & 1,534.26 \\
        & \href{https://basescan.org/address/0xaDDD8ECec572E3BA9975578e1927221eC25Dab50}{\texttt{0xaDDD8ECec572E3BA9975578e1927221eC25Dab50}} & 227.97 & 0.18 & 12.66 & 13.26 & 39.66 \\
        & \href{https://basescan.org/address/0x3f482cA03E6d5FA63884918d5dC379F14cA0aD86}{\texttt{0x3f482cA03E6d5FA63884918d5dC379F14cA0aD86}} & 1,777.25 & 14.27 & 104.54 & 61.42 & 293.66 \\
        & \href{https://basescan.org/address/0xf4Cd829526070C2C0af9E6Dce2C9255130f7ec68}{\texttt{0xf4Cd829526070C2C0af9E6Dce2C9255130f7ec68}} & 139.35 & 0.02 & 9.95 & 1.02 & 39.40 \\
        \midrule
        \multirow{10}{*}{Optimism}
        & \href{https://optimistic.etherscan.io/address/0x7518Ba8b4021D83caFB1e46Fa1250f54fA86b6Ab}{\texttt{0x7518Ba8b4021D83caFB1e46Fa1250f54fA86b6Ab}} & 13,529.19 & 3.75 & 80.05 & 20.25 & 1,346.79 \\
        & \cellcolor{gray!20}\href{https://optimistic.etherscan.io/address/0x8888888885FcA4862619cAFeda9b03768C930012}{\texttt{0x8888888885FcA4862619cAFeda9b03768C930012}} & \cellcolor{gray!20}484.90 & \cellcolor{gray!20}0.02 & \cellcolor{gray!20}4.33 & \cellcolor{gray!20}2.09 & \cellcolor{gray!20}26.72 \\
        & \href{https://optimistic.etherscan.io/address/0x4661f836F07c78a01D5490bFf6e8587B9291C6B8}{\texttt{0x4661f836F07c78a01D5490bFf6e8587B9291C6B8}} & 271.16 & 0.00 & 3.05 & 0.08 & 83.65 \\
        & \href{https://optimistic.etherscan.io/address/0x4db5FBb9246F66bd83661D680F308cc05c5FC3Bf}{\texttt{0x4db5FBb9246F66bd83661D680F308cc05c5FC3Bf}} & 6,849.72 & 1.54 & 77.84 & 16.75 & 989.70 \\
        & \href{https://optimistic.etherscan.io/address/0x874Bd329Adf526A61FeE5766B9E96c9519aeBb1b}{\texttt{0x874Bd329Adf526A61FeE5766B9E96c9519aeBb1b}} & 148.27 & 0.01 & 1.93 & 0.08 & 26.36 \\
        & \href{https://optimistic.etherscan.io/address/0x1dC68046E71429F646Dd6B2d0c02385aE61c2129}{\texttt{0x1dC68046E71429F646Dd6B2d0c02385aE61c2129}} & 5,903.07 & 12.82 & 111.38 & 61.99 & 856.71 \\
        & \cellcolor{gray!20}\href{https://optimistic.etherscan.io/address/0x083CfA7FD187Be983ce5D519fE7ae78357779998}{\texttt{0x083CfA7FD187Be983ce5D519fE7ae78357779998}} & \cellcolor{gray!20}1,083.46 & \cellcolor{gray!20}0.27 & \cellcolor{gray!20}25.80 & \cellcolor{gray!20}16.95 & \cellcolor{gray!20}139.99 \\
        & \href{https://optimistic.etherscan.io/address/0x59c5f1B4FF4fdf04729787EeE84DdBDf787A3033}{\texttt{0x59c5f1B4FF4fdf04729787EeE84DdBDf787A3033}} & 0.17 & 0.00 & 0.00 & 0.00 & 0.06 \\
        & \href{https://optimistic.etherscan.io/address/0x4B1142124c661B48aD7F98d8C7C8db03016B1F88}{\texttt{0x4B1142124c661B48aD7F98d8C7C8db03016B1F88}} & 169.62 & 0.73 & 6.52 & 6.22 & 13.37 \\
        & \href{https://optimistic.etherscan.io/address/0xe3BadF4DCf5908472468e27CF6b438f3F11D9c5D}{\texttt{0xe3BadF4DCf5908472468e27CF6b438f3F11D9c5D}} & 113.64 & 0.00 & 4.37 & 0.35 & 40.95 \\
        \bottomrule
    \end{tabular}
    \end{adjustbox}
    \caption{Overview of the top 10 liquidators across Arbitrum, Base, and Optimism, ranked by the number of liquidations performed, including total realized cost as well as minimum, average, median, and maximum cost per liquidation.}
    \label{tab:liquidator_costs}
\end{table}

\section{Liquidator Price Feeds}
\label{sec:appendix_liquidator_price_feeds}

\tableautorefname{} \ref{tab:feeds_per_liquidator} provides an overview of the Chainlink price feeds targeted by the top 10 liquidators on each chain. These feeds were identified based on the Chainlink price oracles involved in the liquidation transactions executed by each liquidator. Overall, we observe that the highest-ranked liquidators in terms of liquidation count generally target a substantially broader range of price feeds. However, the number of targeted price feeds does not appear to directly correlate with the highest aggregate profit. In addition, all analyzed liquidators target ETH/USD price feeds, while also commonly targeting the native asset of the respective chain, such as ARB/USD on Arbitrum and OP/USD on Optimism. ETH and stablecoin price feeds appear to constitute the most common triggers for liquidation events across all three chains.

\begin{table}[H]
    \centering
    \begin{adjustbox}{max width=\columnwidth}
    \begin{tabular}{l c l}
        \toprule
        \textbf{Chain} & \textbf{Liquidator} & \textbf{Chainlink Price Feeds (Token/USD)} \\
        \midrule
        \multirow{10}{*}{Arbitrum}
        & \cellcolor{gray!20}\href{https://arbiscan.io/address/0xb9516c655831917704A5000bA6f4EB64E816DC2d}{\texttt{0xb9516c655831917704A5000bA6f4EB64E816DC2d}} & \cellcolor{gray!20}AAVE(11), ARB(43), BTC(21), DAI(5), ETH(38), LINK(25), LUSD(1), USDC(93), USDT(26) \\
        & \href{https://arbiscan.io/address/0x001f8151FC6d0a14608B48F1d2a2AeA66c80cF38}{\texttt{0x001f8151FC6d0a14608B48F1d2a2AeA66c80cF38}} & AAVE(22), ARB(35), BTC(26), DAI(1), ETH(42), LINK(26), USDC(94), USDT(32) \\
        & \href{https://arbiscan.io/address/0x48daab9f7Ed6E3184c26D9dDACcc356D52D6237F}{\texttt{0x48daab9f7Ed6E3184c26D9dDACcc356D52D6237F}} & AAVE(7), ARB(62), BTC(5), ETH(5), LINK(21), USDC(59), USDT(23) \\
        & \cellcolor{gray!20}\href{https://arbiscan.io/address/0xd249aB6aaC55a7AD0ceF7fCFD672d47387D7e70F}{\texttt{0xd249aB6aaC55a7AD0ceF7fCFD672d47387D7e70F}} & \cellcolor{gray!20}AAVE(9), ARB(25), BTC(11), DAI(1), ETH(20), LINK(10), USDC(38), USDT(18) \\
        & \cellcolor{gray!20}\href{https://arbiscan.io/address/0x6290c280f1393d33f04d6A59993cbe7d3ECCDFcF}{\texttt{0x6290c280f1393d33f04d6A59993cbe7d3ECCDFcF}} & \cellcolor{gray!20}AAVE(3), ARB(23), BTC(7), ETH(15), LINK(11), USDC(36), USDT(14) \\
        & \href{https://arbiscan.io/address/0x545875b6975eb527726C87B82B5f031ED861De8A}{\texttt{0x545875b6975eb527726C87B82B5f031ED861De8A}} & AAVE(4), ARB(11), BTC(11), DAI(1), ETH(25), LINK(6), USDC(39), USDT(17) \\
        & \cellcolor{gray!20}\href{https://arbiscan.io/address/0x9f83664976aDc0abe9C375151B0E5B36eE7703D7}{\texttt{0x9f83664976aDc0abe9C375151B0E5B36eE7703D7}} & \cellcolor{gray!20}AAVE(2), ARB(35), BTC(1), DAI(1), ETH(13), USDC(40), USDT(11) \\
        & \href{https://arbiscan.io/address/0xa441ed20E5A5c51C067549316B86F84CC988D284}{\texttt{0xa441ed20E5A5c51C067549316B86F84CC988D284}} & AAVE(4), ARB(4), BTC(7), ETH(5), LINK(7), USDC(21), USDT(6) \\
        & \href{https://arbiscan.io/address/0x4044E9E4E9d0ef73026106d097DCdB1d609436A7}{\texttt{0x4044E9E4E9d0ef73026106d097DCdB1d609436A7}} & AAVE(4), ARB(9), BTC(1), ETH(5), LINK(1), USDC(9), USDT(8) \\
        & \href{https://arbiscan.io/address/0x4Ae48f73Abc9F6f7A918ff405584B9314A5df0b8}{\texttt{0x4Ae48f73Abc9F6f7A918ff405584B9314A5df0b8}} & AAVE(4), ARB(3), BTC(1), ETH(6), LINK(6), USDC(13), USDT(4) \\
        \multirow{10}{*}{Base}
        & \href{https://basescan.org/address/0xD251c1325c5d7b29C6219912D8648a3149cDF57B}{\texttt{0xD251c1325c5d7b29C6219912D8648a3149cDF57B}} & AAVE(24), AERO(1), BTC(29), ETH(44), EUR(12), EURC(12), USDC(84), cbBTC(1) \\
        & \cellcolor{gray!20}\href{https://basescan.org/address/0xc89c328609aB58E256Cd2b5aB4F4aF2EFb9fcA33}{\texttt{0xc89c328609aB58E256Cd2b5aB4F4aF2EFb9fcA33}} & \cellcolor{gray!20}AAVE(10), AERO(7), BTC(16), ETH(20), EUR(3), EURC(3), USDC(48) \\
        & \cellcolor{gray!20}\href{https://basescan.org/address/0x3ba19cB44eF72B0B198325E17623BcE10Bae2753}{\texttt{0x3ba19cB44eF72B0B198325E17623BcE10Bae2753}} & \cellcolor{gray!20}AAVE(2), BTC(3), ETH(18), EUR(3), EURC(3), USDC(18) \\
        & \href{https://basescan.org/address/0x964AeE3e4E3BBc7245B33dA097030e95EE408170}{\texttt{0x964AeE3e4E3BBc7245B33dA097030e95EE408170}} & BTC(8), ETH(21), EUR(2), EURC(2), USDC(30) \\
        & \cellcolor{gray!20}\href{https://basescan.org/address/0xec5aCd2dfdf3B2e4dcb955C7eC8B2b74605d8E9c}{\texttt{0xec5aCd2dfdf3B2e4dcb955C7eC8B2b74605d8E9c}} & \cellcolor{gray!20}BTC(3), ETH(9), USDC(12) \\
        & \cellcolor{gray!20}\href{https://basescan.org/address/0x888888887A487f209e31a692B227d8D1ff9070ba}{\texttt{0x888888887A487f209e31a692B227d8D1ff9070ba}} & \cellcolor{gray!20}ETH(21), EUR(1), EURC(1), USDC(20) \\
        & \href{https://basescan.org/address/0x3347277366deCC91d65D2762E792E6BaA471b805}{\texttt{0x3347277366deCC91d65D2762E792E6BaA471b805}} & AAVE(8), BTC(3), ETH(10), EUR(4), EURC(4), USDC(17) \\
        & \href{https://basescan.org/address/0xaDDD8ECec572E3BA9975578e1927221eC25Dab50}{\texttt{0xaDDD8ECec572E3BA9975578e1927221eC25Dab50}} & BTC(6), ETH(12), EUR(3), EURC(3), USDC(15) \\
        & \href{https://basescan.org/address/0x3f482cA03E6d5FA63884918d5dC379F14cA0aD86}{\texttt{0x3f482cA03E6d5FA63884918d5dC379F14cA0aD86}} & BTC(1), ETH(17), EUR(3), EURC(3), USDC(13) \\
        & \href{https://basescan.org/address/0xf4Cd829526070C2C0af9E6Dce2C9255130f7ec68}{\texttt{0xf4Cd829526070C2C0af9E6Dce2C9255130f7ec68}} & AAVE(1), AERO(3), BTC(1), BTC(1), ETH(4), USDC(12), VIRTUAL(4) \\
        \midrule
        \multirow{10}{*}{Optimism}
        & \href{https://optimistic.etherscan.io/address/0x7518Ba8b4021D83caFB1e46Fa1250f54fA86b6Ab}{\texttt{0x7518Ba8b4021D83caFB1e46Fa1250f54fA86b6Ab}} & AAVE(10), BTC(13), DAI(3), ETH(73), LINK(5), OP(98), USDC(123), USDT(12) \\
        & \cellcolor{gray!20}\href{https://optimistic.etherscan.io/address/0x8888888885FcA4862619cAFeda9b03768C930012}{\texttt{0x8888888885FcA4862619cAFeda9b03768C930012}} & \cellcolor{gray!20}AAVE(9), BTC(2), DAI(5), ETH(23), LINK(19), OP(55), SUSD(1), USDC(68), USDT(36) \\
        & \href{https://optimistic.etherscan.io/address/0x4661f836F07c78a01D5490bFf6e8587B9291C6B8}{\texttt{0x4661f836F07c78a01D5490bFf6e8587B9291C6B8}} & AAVE(5), BTC(3), DAI(5), ETH(23), LINK(2), OP(71), SNX(7), SUSD(3), USDC(51), USDT(15) \\
        & \href{https://optimistic.etherscan.io/address/0x4db5FBb9246F66bd83661D680F308cc05c5FC3Bf}{\texttt{0x4db5FBb9246F66bd83661D680F308cc05c5FC3Bf}} & BTC(6), DAI(5), ETH(5), OP(68), USDC(42), USDT(47) \\
        & \href{https://optimistic.etherscan.io/address/0x874Bd329Adf526A61FeE5766B9E96c9519aeBb1b}{\texttt{0x874Bd329Adf526A61FeE5766B9E96c9519aeBb1b}} & DAI(5), ETH(16), OP(67), USDC(39), USDT(22) \\
        & \href{https://optimistic.etherscan.io/address/0x1dC68046E71429F646Dd6B2d0c02385aE61c2129}{\texttt{0x1dC68046E71429F646Dd6B2d0c02385aE61c2129}} & AAVE(4), BTC(1), DAI(4), ETH(3), LINK(6), OP(37), SNX(5), USDC(34), USDT(12) \\
        & \cellcolor{gray!20}\href{https://optimistic.etherscan.io/address/0x083CfA7FD187Be983ce5D519fE7ae78357779998}{\texttt{0x083CfA7FD187Be983ce5D519fE7ae78357779998}} & \cellcolor{gray!20}AAVE(3), BTC(6), DAI(5), ETH(7), OP(21), SNX(6), USDC(25), USDT(9), WBTC(1) \\
        & \href{https://optimistic.etherscan.io/address/0x59c5f1B4FF4fdf04729787EeE84DdBDf787A3033}{\texttt{0x59c5f1B4FF4fdf04729787EeE84DdBDf787A3033}} & AAVE(8), DAI(6), ETH(20), OP(8), USDC(15), USDT(2) \\
        & \href{https://optimistic.etherscan.io/address/0x4B1142124c661B48aD7F98d8C7C8db03016B1F88}{\texttt{0x4B1142124c661B48aD7F98d8C7C8db03016B1F88}} & AAVE(1), ETH(2), LINK(8), OP(17), USDC(18), USDT(6) \\
        & \href{https://optimistic.etherscan.io/address/0xe3BadF4DCf5908472468e27CF6b438f3F11D9c5D}{\texttt{0xe3BadF4DCf5908472468e27CF6b438f3F11D9c5D}} & AAVE(8), BTC(1), LINK(11), OP(6), USDC(16), USDT(10) \\
        \bottomrule
    \end{tabular}
    \end{adjustbox}
    \caption{Overview of the Chainlink price feeds referenced by liquidation transactions for the top 10 liquidators on each chain. Numbers in parentheses denote the number of liquidations executed by the corresponding liquidator that utilized the respective price feed pair. Rows highlighted in gray correspond to liquidators exhibiting speculative behavior.}
    \label{tab:feeds_per_liquidator}
\end{table}

\section{Transaction Distance Optimization}
\label{sec:appendix_transaction_distance_optimization}

\tableautorefname{} \ref{tab:transaction_distance} summarizes the results of our priority fee estimation analysis on Base. Using the \texttt{eth\_feeHistory} RPC method, we estimate priority fees from rolling windows of preceding blocks and compare them against the fees paid by oracle update transactions. The success rate corresponds to the fraction of cases where the estimated fee would still enable same-block inclusion immediately after the oracle update transaction.
Across all analyzed price feeds on Base, our approach achieves high same-block inclusion success rates. For ETH/USD, using the median 50th percentile fee over the previous 37 blocks yields a success rate of 94.48\% with an exact match ratio of 7.64\%.

\begin{table}[H]
    \centering
    \footnotesize
    \begin{tabular}{c r r r}
        \toprule
        \textbf{Price Feed} & \textbf{Fee History (Blocks)} & \textbf{Success Rate (\%)} & \textbf{Exact Match (\%)} \\
        \midrule
        \multirow{2}{*}{BTC / USD} & 49 & \textbf{94.32} & 4.92 \\
        & 1 & 87.22 & \textbf{7.21} \\
        \multirow{2}{*}{ETH / USD} & 37 & \textbf{94.48} & 5.87 \\
        & 1 & 88.99 & \textbf{7.64} \\
        \multirow{2}{*}{AAVE / USD} & 50 & \textbf{94.16} & 7.79 \\
        & 7 & 92.29 & \textbf{13.04} \\
        \bottomrule
    \end{tabular}
    \caption{Success rates and exact match ratios for priority fee estimates based on historical fees, targeting same-block inclusion alongside price oracle update transactions on Base across various price feeds.}
    \label{tab:transaction_distance}
\end{table}

\end{document}